\documentclass[preprint]{jpsj3}

\usepackage{graphicx}
\usepackage{color}

\title{
Theoretical Study of $L$-edge Resonant Inelastic X-ray Scattering in La$_2$CuO$_4$ 
on the Basis of Detailed Electronic Band Structure}

\author{Takuji \textsc{Nomura}\thanks{E-mail address: nomurat@spring8.or.jp} }

\inst{
Japan Atomic Energy Agency, SPring-8, 1-1-1 Kouto, 
Sayo, Hyogo 679-5148, Japan \\}

\recdate{\today}

\abst{
We study theoretically resonant inelastic x-ray scattering (RIXS) at the Cu $L_3$-edge 
in a typical parent compound of high-$T_c$ cuprate superconductors 
La$_2$CuO$_4$ on the basis of a detailed electronic band structure. 
We construct a realistic and precise tight-binding model by employing 
the maximally-localized Wannier functions  
derived from a first-principles electronic structure calculation, 
and then take account of the Coulomb repulsion between $d$ electrons at each Cu site. 
The antiferromagnetic ground state is described within the Hartree-Fock 
approximation, and take account of electron correlations 
in the intermediate states of RIXS within the random-phase approximation (RPA). 
Calculated RIXS spectra agree well with the experimentally observed features 
including low-energy magnon excitation, $d$-$d$ excitations, and charge-transfer excitations, 
over a wide excitation-energy range. 
In particular, we stress the importance of photon polarization dependence: 
the intensity of magnon excitation and the spectral structure of $d$-$d$ excitations 
depend significantly not only on the polarization direction of incident incoming 
photons but also that of outgoing photons. 
It is demonstrated that the single-magnon excitation intensity is maximized 
when the polarization directions of incoming and outgoing photons are perpendicular 
to each other.}

\begin{document}
\maketitle

\section{Introduction}

Resonant inelastic x-ray scattering (RIXS) provides a powerful optical 
method of observing elementary excitations in solids. 
Particularly, recent tremendous progress in energy-momentum resolution 
and theoretical understanding is promoting RIXS to one of major techniques 
of measuring various electronic and magnetic excitations~\cite{Ament2011, Ishii2013}. 
Among RIXS phenomena, RIXS at the transition-metal edges is attracting much 
interest, which is much suitable for elucidating electronic excitations 
in strongly correlated transition-metal compounds 
including the high-$T_c$ cuprate superconductors and various magnetic materials. 
It is notable that RIXS can provide us a variety of electronic excitation spectra, 
depending on which absorption edge of transition metal is utilized. 
At the $K$-edge in the hard x-ray regime, an inner-shell $1s$ electron is promoted 
to conduction $p$ bands. Accompanied with this promotion, 
strongly correlated $d$ electrons near the Fermi level are excited 
to screen the created inner-shell $1s$ hole. 
Thus correlated $d$-electrons are {\it indirectly} excited. 
In this case, electronic excitations should conserve the total spin moment 
of the $d$-electron system, and therefore are restricted only 
to charge-orbital excitation processes or to generation processes of even-number magnons. 
In fact, charge-transfer excitations~\cite{Hill1998,Kim2002}, 
$d$-$d$ orbital excitations~\cite{Ishii2011}, and two-magnon excitations~\cite{Hill2008} 
are observed in transition-metal compounds through $K$-edge RIXS. 
On the other hand, at the $L$-edge in the soft x-ray regime, 
an inner-shell $2p$ electron is promoted to correlated conduction $d$ bands. 
Thus correlated $d$-electrons are {\it directly} excited. 
Since the spin and orbital angular momenta of the $2p$ states 
polarize each other due to strong spin-orbit coupling, 
the $d$ electrons can be excited not only 
in the charge-orbital channel but also in the spin channel. 
In application to high-$T_c$ cuprate superconductors, 
in which the spin moment is induced on the $d_{x^2-y^2}$ orbital, 
it had been believed that spin-flip scattering is not allowed 
in $L$-edge RIXS~\cite{DeGroot1998, VanVeenendaal2006}. 
Actually, more recently Ament and collaborators verified 
that spin-flip excitation is indeed allowed when the spin moments point parallel 
to the basal plane~\cite{Ament2009}. 
In fact, magnon excitation is clearly observed in copper oxides 
where the spin moments are aligned along the basal 
plane~\cite{Braicovich2010a,Braicovich2010b,Guarise2010,Schlappa2012,Dean2012}. 
Magnon excitation is observed persistently even in doped metallic 
cuprates~\cite{LeTacon2011,Jia2014,Ishii2014,Lee2014}, 
and $L$-edge RIXS promises to be a new technique of elucidating 
the pairing glue in high-$T_c$ superconductivity. 
The observed energy dispersion of the magnon peak agrees well 
with neutron scattering experiments~\cite{Braicovich2010a,Braicovich2010b,
Guarise2010,LeTacon2011,Dean2012}. 

In our present work, we discuss theoretically RIXS at the transition-metal $L$-edge. 
Intensive theoretical studies have been performed so far 
to analyze the $L$-edge RIXS in strongly correlated transition-metal 
compounds~\cite{Tanaka1993,Tsutsui2006, Ament2009, 
Haverkort2010, Kaneshita2011, Igarashi2012}. 
In most of precedent studies, simplified effective models 
such as impurity Anderson models, finite-size cluster models, 
or Heisenberg models have been adopted to describe the electronic structure 
or the magnetic ground state, and the energy window and the number 
of momentum points allowed for calculation are inevitably restricted. 
Heisenberg antiferromagnetic superexchange couplings are often treated 
as tunable parameters to fit to experimental data, 
and microscopic quantitative grounds for the values of coupling parameters are lacking. 
Therefore, it will be meaningful to develop another approximate but useful calculations 
without the above drawbacks, for analyzing experimental data in detail. 

In the present study, we present a microscopic theoretical formulation of RIXS 
at the $L$-edge and apply it to a typical parent compound of high-$T_c$ 
superconductors La$_2$CuO$_4$. 
To describe the electronic structure of La$_2$CuO$_4$ precisely, 
we use maximally localized Wannier functions (MLWF) derived 
from first-principles electronic structure calculation~\cite{Mostofi2008}, 
and determine the antiferromagnetic ground state within the Hartree-Fock (HF) approximation. 
Electron correlations in the intermediate states are treated 
within the random-phase approximation (RPA). 
Our approach is based on perturbation expansion in Coulomb interaction, 
and a natural extension of our previous theoretical formulation 
of $K$-edge RIXS~\cite{Nomura2005, Nomura2014}. 
It is approximate one but applicable to realistic and complex electronic structures 
such as multi-orbital systems. 
Similar formulation is already developed in Ref.~\ref{Ref:Igarashi2013}. 

The article is constructed in the following way: 
In \S~\ref{Sc:Formulation}, we present our Hamiltonian and theoretical formulation 
to calculate RIXS spectra. 
In \S~\ref{Sc:Results}, we present calculated RIXS spectra 
of single-magnon and $d$-$d$ excitations, and their dependences 
on polarization direction and scattering angles, 
and compare some of them with experimental data. 
In particular, we shall see the intensity of magnon excitation and the spectral structure 
of $d$-$d$ excitations depend significantly not only on the polarization direction 
of incident photons but also that of outgoing photons. 
In \S~\ref{Sc:Discussions}, some discussions and remarks 
on our formulation and results are given. 
In \S~\ref{Sc:Conclusion}, the article is concluded with brief comments. 

\section{Formulation of RIXS}
\label{Sc:Formulation}

\subsection{Hamiltonian}
\label{Sc:Hamiltonian}
To discuss the RIXS process microscopically, 
we use the following form of Hamiltonian: 
\begin{equation}
H = H_{n.f.} + H_{2p} + H_{2p-d} + H_x, 
\end{equation}
where $H_{n.f.}$ describes the correlated electrons near the Fermi level, 
and $H_{2p}$ and $H_x$ describe the inner-shell $2p$ electrons 
and the dipole transition by x-rays, respectively. 
$H_{2p-d}$ is the Coulomb interaction between the $2p$ 
and transition-metal $d$ electrons 
(Cu-3$d$ electrons in the case of La$_2$CuO$_4$). 
We present details of each term in the following. 

\begin{figure}
\begin{center}
\includegraphics[width=100mm]{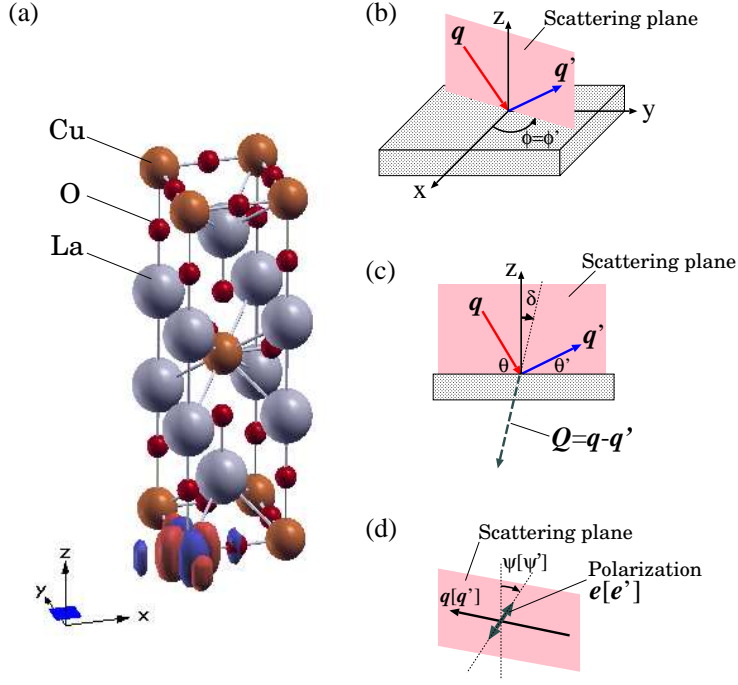}
\end{center}
\caption{
(Color online) 
(a) Crystal structure of La$_2$CuO$_4$ and MLWF for the Cu-$d_{x^2-y^2}$ state. 
(b), (c) and (d) Definition of the scattering geometry. 
$\mib{q}$ and $\mib{q}'$ are the momentum vectors 
of the incoming and outgoing photons, respectively. 
$\theta$,  $\phi$ and $\psi$ [$\theta$',  $\phi$' and $\psi$'] 
are the Bragg, azimuthal and polarization angles, respectively, 
for incoming [outgoing] photons. 
In (c), the angle between transferred momentum $\mib{Q}$ 
and the $z$-axis is approximately given by $\delta \approx (\theta - \theta')/2$. 
In (d), $\mib{e}$ and $\mib{e}'$ are the polarization vectors 
of the incoming and outgoing photons, respectively. 
The polarization angle is measured with respect to the scattering plane, 
i.e., $\psi=0$ [$\psi=\pi/2$] means that the polarization direction 
is parallel [perpendicular] to the scattering plane.}
\label{Fig:geometry}
\end{figure}

To construct the Hamiltonian part $H_{n.f.}$ for La$_2$CuO$_4$, 
firstly we perform first-principles band structure calculation 
assuming the paramagnetic state~\cite{Blaha2013}. 
To express the electronic orbital bases and scattering geometry, 
we take the coordinate system where the principal axes 
of nearest-neighbor Cu-O bonds are parallel along the cartesian axes 
(see Fig~\ref{Fig:geometry}(a)). 
Then we perform tight-binding fitting to the obtained energy bands 
near the Fermi level by using the {\tt wannier90} code~\cite{Mostofi2008}, 
where we take five $d$ orbitals at each Cu, and three $p$ orbitals at each O site. 
Thus we include 17 MLWF's in the unit cell, since there are one Cu and four O sites in the unit cell. 
The orbital bases are defined in terms of the coordinate axes in Fig.~\ref{Fig:geometry}(a), 
and the spin states are specified with respect to the $z$-axis. 
Thus we obtain a tight-binding model to fit the 17 bands near the Fermi level 
(See Appendix A for details). 
Adding the on-site Coulomb interaction part, 
we have the Hubbard-type Hamiltonian $H_{n.f.}$ in the following form: 
\begin{equation}
H_{n.f.} = \sum_{ii'} \sum_{\ell}^{@\mib{r}_i} \sum_{\ell'}^{@\mib{r}_{i'}} 
\sum_\sigma t_{\ell,\ell'}({\mib r}_i-\mib{r}_{i'}) 
a_{i \ell \sigma}^{\dag} a_{i' \ell' \sigma} + \frac{1}{2} \sum_i^{\rm t.m.} \sum_{\ell_1 \sim \ell_4}^{@\mib{r}_i} 
\sum_{\sigma\sigma'} I_{\ell_1, \ell_2; \ell_3, \ell_4}(\mib{r}_i) 
a_{i \ell_1 \sigma}^{\dag} a_{i \ell_2 \sigma'}^{\dag} 
a_{i \ell_3 \sigma'} a_{i \ell_4 \sigma}, 
\end{equation}
where $a_{i \ell \sigma}^{\dag}$ and $a_{i \ell \sigma}$ 
are the electron creation and annihilation operators 
for orbital $\ell$ with spin $\sigma$ at site $i$. 
Throughout the present study, we always work 
with the electron representation not with the hole representation. 
`$\rm{t.m.}$' in the summation with respect to $i$ means summing 
only over transition-metal sites. 
$I_{\ell_1,\ell_2;\ell_3,\ell_4}(\mib{r}_i) \equiv I_{\ell_1,\ell_2;\ell_3,\ell_4}$ 
is the on-site Coulomb integral at transition-metal (i.e., Cu) sites. 
In the summation with respect to $\ell$, `$@\mib{r}_i$' at the top 
means orbital $\ell$ should lie on the site $\mib{r}_i$. 
One-particle energy at orbital $\ell$ is given 
by $\varepsilon_{\ell} \equiv t_{\ell,\ell}(\mib{r}=0)$. 
We modify the one-particle energy $\varepsilon_{\ell}$ for Cu-$d$ orbitals, 
to obtain a realistic level scheme of the local Cu-$d$ orbitals and $d$-$d$ 
excitation energies, as explained in Appendix A. 
Hereafter we use the following convention for later discussions: 
if $\ell$ denotes a $d$ orbital at a Cu site (e.g., $\ell=xy$), 
then $a_{i \ell \sigma} \equiv d_{i \ell \sigma} = d_{i \zeta} $, 
if $\ell$ denotes a $p$ orbital (e.g., $\ell=x$) at an O site, 
then $a_{i \ell \sigma} \equiv p_{i \ell \sigma}$, and so on. 
We expect that confusion between Cu-$p$ and O-$p$ orbitals will not occur, 
since the O-$p$ orbitals do not appear explicitly in the following discussions. 
Here we introduce the values of on-site Coulomb interaction $I_{\ell_1,\ell_2;\ell_3,\ell_4}$ 
at each Cu site in the form of Slater-Condon integrals 
(see Ref.~\ref{Ref:Condon1959} for the definition 
of Slater-Condon integrals and their relation 
to $I_{\ell_1,\ell_2;\ell_3,\ell_4}$): $F^0(d,d) = 6$ eV, 
$F^2(d,d) = 11.5$ eV, $F^4(d,d) = 7.4$ eV. 
The values of $F^2(d,d)$ and $F^4(d,d)$ are taken 
from Ref.~\ref{Ref:Czyzyk1994}, 
and $F^0(d,d)$ is determined to reproduce the insulating gap 
about 2.2 eV~\cite{Ginder1988}. 
Our choice of these Coulomb integrals corresponds 
approximately to $U \sim 7.5$ eV, $U' \sim 5$-6 eV, 
and $J \sim 0.6$-1.2 eV, where $U$, $U'$ and $J$ 
are the intra-orbital and inter-orbital Coulomb repulsions 
and the Hund's coupling, respectively. 
For $H_{n.f.}$,  we determine the antiferromagnetic ground state 
with the spin moments $\mib{m} \parallel [110]$ 
within the HF approximation (see Appendix A about details of HF calculation). 

For the $2p$ electrons, we assume completely localized $2p$ orbitals 
at each transition-metal site (Cu site in the present case of La$_2$CuO$_4$): 
\begin{eqnarray}
H_{2p} = \sum_i^{\rm t.m.} \biggl[ \sum_{m, \sigma}^{@\mib{r}_i} 
\varepsilon_{2p}(\mib{r}_i) p_{i m \sigma}^{\dag} p_{i m \sigma} + \sum_{mm'}^{@\mib{r}_i} 
\sum_{\sigma \sigma'} \xi_{2p}(\mib{r}_i) \mib{l}_{mm'} \cdot \mib{s}_{\sigma\sigma'} 
p_{i m \sigma}^{\dag} p_{i m' \sigma'} \biggr], 
\label{eq:2phamiltonian}
\end{eqnarray}
where $\varepsilon_{2p}(\mib{r}_i) \equiv \varepsilon_{2p}$ is the one-particle energy 
of the $2p$ state, $p_{i m \sigma}^{\dag}$ and $p_{i m \sigma}$ are the creation 
and annihilation operators of $2p$ electrons with spin $\sigma$ and angular momentum $m$ 
($m=-1,0,1$) at transition-metal site $i$, respectively. 
The second term of the right-hand side of eq.~(\ref{eq:2phamiltonian}) is the spin-orbit coupling. 
$\mib{l}_{mm'}$ are the matrix elements of orbital angular momentum, and 
$\mib{s}_{\sigma\sigma'}$ are related to the Pauli matrices as $\mib{s}=\mib{\sigma}/2$. 
$H_{2p}$ is straightforwardly diagonalized, and the $2p$ eigenstates are characterized by the total 
angular momentum $J$ and its $z$ component $M$, 
where $J=1/2, 3/2$ and $M=-J, -J+1, ... , J-1, J$. 
The energy levels consist of two kinds of levels as a result from splitting due to the spin-orbit coupling:  
$\varepsilon_{{2p}_{J=1/2}} = \varepsilon_{2p} - \xi_{2p}$ 
and $\varepsilon_{2p_{J=3/2}} = \varepsilon_{2p} + \xi_{2p}/2 $. 
The former (latter) is two-fold (four-fold) degenerated. 
In the present case of Cu, we take $\varepsilon_{2p} = -936$ eV with respect to the Fermi level, 
and  $\xi_{2p} = 13.3$ eV, which causes about 20 eV energy splitting 
between the $L_2$ and $L_3$ edges. 
The diagonalized eigenstates are expressed as 
\begin{equation}
| J, M \rangle = \sum_{m = 1,0,-1} \sum_{\sigma = \uparrow, \downarrow} 
| m, \sigma  \rangle u_{m\sigma; JM}. 
\end{equation}
Specifically, for the $J=1/2$ doublet, 
\begin{eqnarray}
| J=1/2, M=1/2  \rangle &=& - \sqrt{\frac{1}{3}} | m=0, \uparrow  \rangle 
+ \sqrt{\frac{2}{3}} | m=1, \downarrow  \rangle  \\ 
| J=1/2, M=-1/2 \rangle &=& \sqrt{\frac{2}{3}} | m=-1, \uparrow \rangle 
- \sqrt{\frac{1}{3}} | m=0, \downarrow  \rangle, 
\end{eqnarray}
and for the $J=3/2$ quartet, 
\begin{eqnarray}
| J=3/2, M=3/2  \rangle &=& | m=1,  \uparrow  \rangle \\ 
| J=3/2, M=1/2 \rangle &=& \sqrt{\frac{2}{3}} | m=0, \uparrow  \rangle 
+ \sqrt{\frac{1}{3}} | m=1,  \downarrow  \rangle  \\
| J=3/2, M=-1/2  \rangle &=& \sqrt{\frac{1}{3}} | m=-1,  \uparrow  \rangle  
+ \sqrt{\frac{2}{3}} | m=0,  \downarrow  \rangle  \\ 
| J=3/2, M=-3/2 \rangle &=& | m=-1, \downarrow  \rangle. 
\end{eqnarray}

$H_{2p-d}$ is given by
\begin{equation}
H_{2p-d} = \sum_i^{\rm t.m.} \sum_{\ell\ell'}^{@\mib{r}_i} 
\sum_{mm'}^{@\mib{r}_i} \sum_{\sigma_1 \sim \sigma_4} 
V_{2p-d}( \mib{r}_i; m\sigma_1, \ell\sigma_2; \ell'\sigma_3, m'\sigma_4) 
p_{i m \sigma_1}^{\dag} d_{i \ell \sigma_2}^{\dag} d_{i \ell' \sigma_3} p_{i m' \sigma_4}, 
\end{equation}
where $V_{2p-d}(\mib{r}_i; \cdots )$ is the on-site Coulomb interaction between 
the $2p$ and $d$ orbitals at transition-metal site $\mib{r}_i$. 
$\ell$ and $\ell'$ denote the $d$ orbitals: $\ell, \ell' = xy, yz, xz, x^2-y^2, 3z^2-r^2$. 
For convenience in later discussions, we define the following matrix elements: 
\begin{eqnarray}
V_{2p-d}( \mib{r}_i; JM, \zeta; \zeta', J'M') \equiv \sum_{mm'} \sum_{\sigma_1\sigma_2} 
u_{m \sigma_1; JM}^* V_{2p-d}( \mib{r}_i; m\sigma_1, \ell\sigma; \ell'\sigma', m'\sigma_2) 
u_{m' \sigma_2; J'M'}. 
\end{eqnarray}
For numerical calculations on La$_2$CuO$_4$, 
we take $F^0(p,d) = F^2(p,d) = 1$ eV at Cu sites, but this interaction is almost 
irrelevant to spectral weights for the case of La$_2$CuO$_4$ as we see later. 

$H_x$ describes resonant $2p$-$d$ dipole transition induced by x-rays: 
\begin{equation}
H_x = \sum_i^{\rm t.m.u} \frac{1}{N} \sum_{\mib{k},\mib{q}} \sum_{\ell, m}^{@\mib{r}_i} \sum_{\sigma} 
w_{\ell, m}(\mib{r}_i; {\mib q}, {\mib e}) \alpha_{\mib{q}\mib{e}} 
d_{{\mib k}+{\mib q} \ell \sigma}^{\dag} p_{{\mib k} m \sigma} + h.c., 
\end{equation}
where $N$ is the number of unit cells, the summation in $i$ with `t.m.u' means 
it is restricted to transition-metal sites in the unit cell. 
$d_{{\mib k} \ell \sigma}^{\dag}$ [$p_{{\mib k} m \sigma}$] 
is the creation [annihilation] operator of transition-metal $d_\ell$ [$2p_m$] electrons 
in the momentum representation, and $\alpha_{\mib{q}\mib{e}}$ 
is the annihilation operator of a photon with momentum $\mib{q}$ and polarization $\mib{e}$. 
We assume the matrix elements of  $w_{\ell, m}(\mib{r}_i; {\mib q},{\mib e})$ 
are given in the form: 
\begin{eqnarray}
w_{\ell, m}(\mib{r}_i ; {\mib q},{\mib e}) &=& - \frac{e}{m_e}\sqrt{\frac{2\pi}{|\mib{q}|}} 
e^{i\mib{q} \cdot \mib{r}_i} \langle d_{\ell} | \mib{e} \cdot \mib{p}| 2p_m \rangle \nonumber\\ 
&=& - \frac{e}{m_e} \sqrt{\frac{2\pi}{|\mib{q}|}} 
e^{i\mib{q} \cdot \mib{r}_i} i m_e (\varepsilon_{d_\ell} - \varepsilon_{2p_m}) 
\langle d_{\ell} | \mib{e} \cdot \mib{r}| 2p_m \rangle \nonumber \\ 
& \approx & - i e \sqrt{ 2\pi |\mib{q}|} e^{i\mib{q} \cdot \mib{r}_i} 
\langle d_{\ell} | \mib{e} \cdot \mib{r}| 2p_m \rangle 
\end{eqnarray}
in natural units ($c = \hbar = 1$), $m_e$ and $e$ are the mass and charge of an electron, respectively. 
$\mib{r}_i$ is the transition-metal site where the x-rays are absorbed or emitted, 
and we assume that $w_{\ell, m}(\mib{r}_i ; {\mib q},{\mib e})$ does not vanish 
only when both of the orbitals $2p_m$ and $d_{\ell}$ lie on the site $\mib{r}_i$. 
$\langle d_{\ell} | \mib{e} \cdot \mib{r}| 2p_m \rangle$ can be calculated using atomic wave functions. 
For convenience in later discussions, we define 
\begin{equation}
w_{\zeta, JM}(\mib{r}_i; {\mib q},{\mib e}) \equiv w_{\ell \sigma, JM}(\mib{r}_i; {\mib q},{\mib e}) 
\equiv \sum_{m = 1,0,-1} w_{\ell, m}(\mib{r}_i; {\mib q},{\mib e}) u_{m\sigma; JM}, 
\end{equation}
where $\zeta$ is orbital-spin unified index: $\zeta \equiv (\ell,\sigma)$. 

\subsection{RIXS intensity}
\label{Sc:RIXSintensity}
\begin{figure}
\begin{center}
\includegraphics[width=90mm]{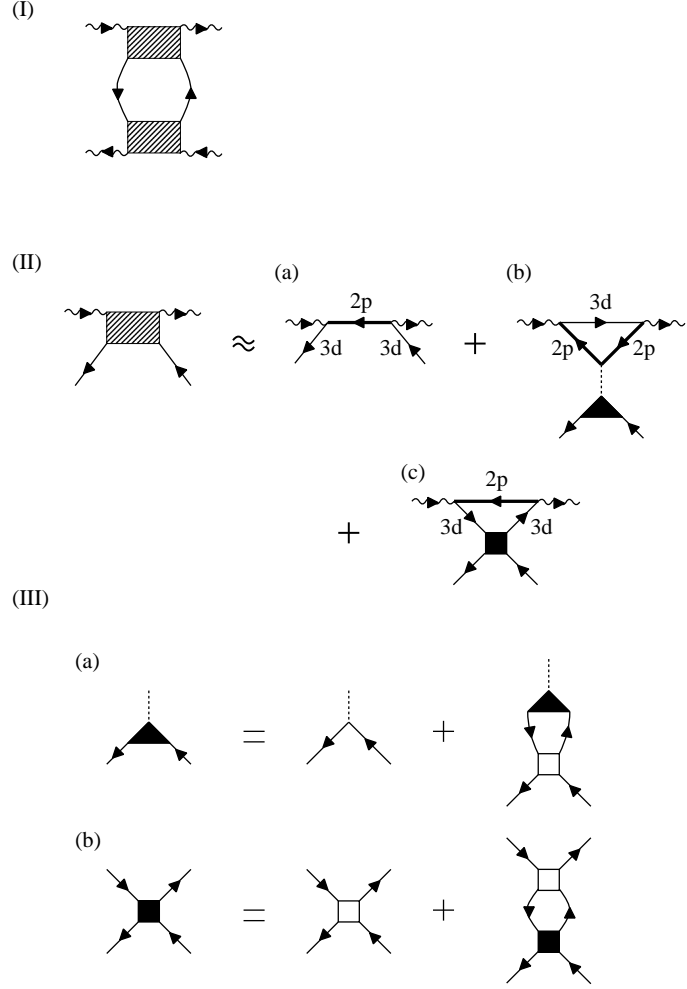}
\end{center}
\caption{
(I) RIXS intensity represented within the Keldysh perturbative formulation. 
The wavy lines and shaded rectangular represent 
the photon propagators and electron scattering vertex function $F(\mib{r}_i; q, q')$, respectively. 
A pair of oriented solid lines represent the off-diagonal elements of the Keldysh Green's function, 
and connect the upper normally-time-ordered and lower reversely-time-ordered branches. 
(II) Approximate expansion for the scattering vertex function $F(\mib{r}_i; k_1; q, q')$: 
(a) $F^{(0)}(\mib{r}_i; \mib{k}_1; q, q')$ for `0th-order process' (fluorescence-like), 
(b) $F^{(p)}(\mib{r}_i; q, q')$ for `$p$-process', (c) $F^{(d)}(\mib{r}_i; q, q')$ for `$d$-process'. 
The filled triangle and square are the three-point and four-point vertex functions 
to be renormalized by electron correlations, respectively. 
In (b), the dashed line represents the core-hole potential $V_{2p-d}$. 
Thick solid lines represent the propagator of the inner-shell $2p$ electrons. 
(III) RPA diagrams for the three-point and four-point vertex functions 
$\Lambda_{\zeta_2,\zeta_4; \zeta_3,\zeta_1}(Q)$ 
and $\Gamma_{\zeta_2,\zeta_4; \zeta_3,\zeta_1}(Q)$ ((a) and (b), respectively), 
where empty squares represent the antisymmetrized bare Coulomb interaction 
$\Gamma_{\zeta_2,\zeta_4; \zeta_3,\zeta_1}^{(0)}$ 
among the $d$ electrons at transition-metal sites.}
\label{Fig:diagrams}
\end{figure}

RIXS intensity can be obtained by calculating the number of photons generated 
in different states from the incident-photon state per unit time. 
To do this, we employ the Keldysh perturbation theory as in Ref.~\ref{Ref:Nozieres1974}. 
The present formulation of $L$-edge RIXS is a straightforward extension of our previous 
formulation on $K$-edge RIXS~\cite{Nomura2005,Nomura2014}. 
The RIXS intensity is generally expressed by the diagram (I) in Fig.~\ref{Fig:diagrams}, 
if assuming that only a single electron-hole pair remains in the final state. 
The analytic expression of RIXS intensity is obtained from the diagram (I) 
of Fig.~\ref{Fig:diagrams} as: 
\begin{eqnarray}
W(q, \mib{e}; q', \mib{e}') &=& \frac{1}{N} \sum_{\mib{k}_1} \int_{-\infty}^{\infty} 
\frac{d \omega_1}{2\pi} \sum_{a_1 a_2} G_{a_1}^+(k_1) G_{a_2}^-(k_1+Q) 
\biggl| \sum_i^{\rm t.m.u.} \sum_{\zeta\zeta'}^{@\mib{r}_i} \sum_{JJ'} 
\sum_{M=-J}^{J} \sum_{M'=-J'}^{J'} \nonumber \\ 
&& w_{\zeta, JM}(\mib{r}_i; \mib{q}, \mib{e}) w_{\zeta', J'M'}^*(\mib{r}_i; \mib{q}', \mib{e}') 
F_{\zeta JM, \zeta' J'M'; a_1, a_2}(\mib{r}_i; k_1; q, q') \biggr|^2, \label{Eq:W}
\end{eqnarray}
where $G_a^\pm(k)$ is the Keldysh Green's function~\cite{Keldysh1965}, 
$a_{1,2}$ are indices for the diagonalized bands, 
$\zeta  = (\ell,\sigma)$, $\zeta' = (\ell',\sigma')$, 
$\sum_{\zeta\zeta'} = \sum_{\ell\ell'}\sum_{\sigma\sigma'}$, 
and $k_1 = (\omega_1, \mib{k}_1)$. 
$q$ and $q'$ are the four-momenta of the incoming and outgoing photons, respectively: 
$q = (\omega, \mib{q})$, $q' = (\omega', \mib{q}')$. 
$\mib{e}$ and $\mib{e}$' are the unit vectors pointing along the polarization direction 
of the absorbed and emitted photons, respectively. 
$Q$ is the energy and momentum loss of the photon: 
$Q = q-q' = (\omega-\omega', \mib{q} - \mib{q}') \equiv (\Omega, \mib{Q})$. 
$F_{\zeta JM, \zeta' J'M'; a_1, a_2}(\mib{r}_i; k_1; q, q')$ 
is the scattering vertex function expressed 
using only the ordinary causal electron Green's functions and electron-electron interaction. 
At this stage, we omit $\omega_1$ dependence 
of $F_{\zeta JM, \zeta' J'M'; a_1, a_2}(\mib{r}_i; k_1; q, q')$, i.e., 
$F_{\zeta JM, \zeta' J'M'; a_1, a_2}(\mib{r}_i; k_1; q, q') = 
F_{\zeta JM, \zeta' J'M'; a_1, a_2}(\mib{r}_i; \mib{k}_1; q, q')$, 
because it is justified within the following approximation 
for $F_{\zeta JM, \zeta' J'M'; a_1, a_2}(\mib{r}_i; k_1; q, q')$. 
Within the HF approximation, the Green's functions $G_a^\pm(k)$ are given by 
\begin{eqnarray}
G_a^+(k_1) &=& 2 \pi i n_a(\mib{k}_1) \delta(\omega_1-E_a(\mib{k}_1)), \label{Eq:G+}\\
G_a^-(k_1) &=& -2 \pi i [1-n_a(\mib{k}_1)] \delta(\omega_1-E_a(\mib{k}_1)), \label{Eq:G-}
\end{eqnarray}
where $E_a(\mib{k}_1)$ is the energy of diagonalized band $a$, 
and $n_a(\mib{k}_1)$ is the electron occupation density i.e., the Fermi distribution function, 
at momentum $\mib{k}_1$ in band $a$: $n_a(\mib{k}_1) = 1/(\exp [E_a(\mib{k}_1)/T] + 1)$. 
Substituting eqs.~(\ref{Eq:G+}) and (\ref{Eq:G-}) into eq.~(\ref{Eq:W}), we have 
\begin{eqnarray}
W(q, \mib{e}; q', \mib{e}') &=& \frac{2 \pi}{N} \sum_{\mib{k}_1} \sum_{a_1 a_2} 
n_{a_1}(\mib{k}_1) [1-n_{a_2}(\mib{k}_1 + \mib{Q}) ]
\delta (\Omega + E_{a_1}(\mib{k}_1)- E_{a_2}(\mib{k}_1 + \mib{Q})) \nonumber \\
&& \times \biggl| \sum_i^{\rm t.m.u.} \sum_{\zeta\zeta'}^{@\mib{r}_i}
\sum_{JJ'} \sum_{M=-J}^{J} \sum_{M'=-J'}^{J'} 
w_{\zeta, JM}(\mib{r}_i; \mib{q}, \mib{e}) 
w_{\zeta', J'M'}^*(\mib{r}_i; \mib{q}', \mib{e}')  \nonumber \\
&& \times F_{\zeta JM, \zeta' J'M'; a_1, a_2}(\mib{r}_i; \mib{k}_1; q, q') \biggr|^2. \label{Eq:W2}
\end{eqnarray}
For calculation of $F_{\zeta JM, \zeta' J'M'; a_1, a_2}(\mib{r}_i; \mib{k}_1; q, q')$, 
we use perturbation expansion with respect to electron-electron interactions. 
There are three possible major contributions 
to $F_{\zeta JM, \zeta' J'M'; a_1, a_2}(\mib{r}_i; \mib{k}_1; q, q')$. 
The first is the zeroth-order term represented by the diagram (II)-(a) 
in Fig.~\ref{Fig:diagrams}. 
This diagram presents the main contribution from fluorescence processes, 
where a valence $d$ electron goes down to the $2p$ state without interacting with other electrons. 
We refer to this contribution as `0th-order process'.
The second originates from the screening of the $2p$ core hole. 
Within the Born approximation with respect to the core-hole potential $V_{2p-d}$, 
this process is expressed by the diagram (II)-(b) in Fig.~\ref{Fig:diagrams}. 
We refer to this contribution as `$p$-scattering process' or `$p$-process'. 
The third describes the scattering processes involving both 
the excited conduction $d$ electron and valence $d$ electrons. 
This contribution is expressed by the diagram (II)-(c) in Fig.~\ref{Fig:diagrams}. 
We refer to this contribution as `$d$-scattering process' or `$d$-process'. 
Of course, in higher-order contributions, more complex diagrams can appear, 
which cannot simply be classified to `$p$-process' or `$d$-process'. 
Nevertheless, this classification turns out to be convenient 
for microscopic analysis of RIXS spectra. 
Thus, we obtain the following approximate expression 
for the scattering vertex function: 
\begin{eqnarray}
F_{\zeta JM, \zeta' J'M'; a_1, a_2}(\mib{r}_i; \mib{k}_1; q, q') 
&=& F^{(0)}_{\zeta JM, \zeta' J'M'; a_1, a_2}(\mib{r}_i; \mib{k}_1; q, q') \nonumber \\
&& - \sum_{\zeta_1\zeta_2} u_{\zeta_2, a_2}^*(\mib{k}_1+\mib{Q}) u_{\zeta_1, a_1}(\mib{k}_1) 
[ F^{(p)}_{\zeta JM, \zeta' J'M'; \zeta_1, \zeta_2}(\mib{r}_i; q, q') \nonumber \\
&& + F^{(d)}_{\zeta JM, \zeta' J'M'; \zeta_1, \zeta_2}(\mib{r}_i; q, q') ], 
\label{Eq:F}
\end{eqnarray}
where $u_{\zeta, a}(\mib{k})$ is the diagonalization matrix 
of the HF Hamiltonian given by eq.~(\ref{Eq:HMF}). $\zeta_n$ is orbital-spin unified index: 
$\zeta_n = (\ell_n, \sigma_n)$, and $\sum_{\zeta_n} = \sum_{\ell_n}\sum_{\sigma_n} $, 
where $\ell_n$ represents $d$ orbitals at transition-metal sites. 
For the antiferromagnetic ordered state, the energy bands are doubly folded. 
Therefore, in addition to $\zeta_n$, a two-valued index is necessary 
to specify which of doubled bands each state is on. 
But we suppress it for simplicity of notation. 
Contributions from the above three processes are given by 
\begin{equation}
F^{(0)}_{ \zeta JM, \zeta' J'M'; a_1, a_2}(\mib{r}_i; \mib{k}_1; q, q') = 
\delta_{JJ'}\delta_{MM'} 
\frac{u_{d_{\zeta}(i), a_2}^*(\mib{k}_1+\mib{Q}) u_{d_{\zeta'}(i), a_1}(\mib{k}_1)}
{\omega + \tilde{\varepsilon}_{2p_J}(\mib{r}_i) - E_{a_2}(\mib{k}_1+\mib{Q})}, 
\end{equation}
\begin{eqnarray}
F^{(p)}_{ \zeta JM, \zeta' J'M'; \zeta_1, \zeta_2}(\mib{r}_i; q, q') &=& 
\sum_{\zeta_3\zeta_4}^{@\mib{r}_i} V_{2p-d}(\mib{r}_i; JM, \zeta_3; \zeta_4, J'M') 
\Lambda_{\zeta_2,\zeta_4; \zeta_3,\zeta_1}(Q) \nonumber \\
&& \times \sum_a \frac{1}{N} \sum_{\mib{k}} [1-n_a(\mib{k})] \nonumber\\
&& \times \frac{u_{d_{\zeta}(i), a}^*(\mib{k}) u_{d_{\zeta'}(i), a}(\mib{k})}
{[\omega + \tilde{\varepsilon}_{2p_J}(\mib{r}_i) - E_a(\mib{k})]
[\omega' + \tilde{\varepsilon}_{2p_{J'}}(\mib{r}_i) - E_a(\mib{k})]}, 
\end{eqnarray}
\begin{eqnarray}
F^{(d)}_{ \zeta JM, \zeta' J'M'; \zeta_1, \zeta_2}(\mib{r}_i; q, q') = \delta_{JJ'}\delta_{MM'}
\sum_{\zeta_3\zeta_4} \Gamma_{\zeta_2,\zeta_4; \zeta_3,\zeta_1}(Q)
\sum_{a_3 a_4} \frac{1}{N} \sum_{\mib{k}} [1-n_{a_3}(\mib{k}+\mib{Q})] \nonumber \\
\times \frac{u_{d_{\zeta}(i), a_3}^*(\mib{k}+\mib{Q}) u_{\zeta_3, a_3}(\mib{k}+\mib{Q}) 
u_{\zeta_4, a_4}^*(\mib{k}) u_{d_{\zeta'}(i), a_4}(\mib{k})} 
{\omega + \tilde{\varepsilon}_{2p_J}(\mib{r}_i) - E_{a_3}(\mib{k}+\mib{Q})} \nonumber\\
\times \biggl( \frac{1-n_{a_4}(\mib{k})}{\omega' + \tilde{\varepsilon}_{2p_{J'}}(\mib{r}_i) - E_{a_4}(\mib{k})} 
- \frac{n_{a_4}(\mib{k})}{\Omega+E_{a_4}(\mib{k}) - E_{a_3}(\mib{k}+\mib{Q}) + i\epsilon} \biggr),  
\nonumber \\
\label{Eq:Fd}
\end{eqnarray}
where $\Lambda_{\zeta_2, \zeta_4; \zeta_3, \zeta_1}(Q)$ and 
$\Gamma_{\zeta_2,\zeta_4; \zeta_3,\zeta_1}(Q)$ are the three-point and 
four-point vertex functions, which are represented by the filled triangle 
and square in Fig.~\ref{Fig:diagrams} (II) (b) and (c), respectively. 
Index $d_\zeta(i) = d_{\ell\sigma}(i)$ denotes the $d_\ell$ state 
at transition-metal site $\mib{r}_i$ with spin $\sigma$. 
$\tilde{\varepsilon}_{2p_J}(\mib{r}_i) \equiv \varepsilon_{2p_J}(\mib{r}_i) + i \Gamma_{2p} $, 
where $\Gamma_{2p}$ is the damping rate of the $2p$ core hole 
and set to 0.3 eV in the present study. 
Summations in $i$ with `t.m.u.' at the top means that $\mib{r}_i$ 
should be restricted only to transition-metal sites in the unit cell. 
$\epsilon$ in eq.~(\ref{Eq:Fd}) ensures the causality, 
i.e., that an electron-hole pair is created only after the resonant transition of a $2p$ electron to $d$ states. $\epsilon$ 
has the physical meaning of the damping of the excited electron-hole pair. 
In our numerical calculations, we set $\epsilon = 20$ meV. 

The vertex functions introduced above are renormalized by electron correlations. 
We take account of electron correlations within RPA. 
RPA for $\Lambda_{\zeta_2,\zeta_4; \zeta_3,\zeta_1}(Q)$ and 
$\Gamma_{\zeta_2,\zeta_4; \zeta_3,\zeta_1}(Q)$ 
is represented diagrammatically in Fig.~\ref{Fig:diagrams} (III) (a) and (b), respectively. 
The analytic expressions for these diagrams are 
\begin{eqnarray}
\Lambda_{\zeta_2,\zeta_4; \zeta_3,\zeta_1}(Q) = \delta_{\zeta_2\zeta_3}\delta_{\zeta_4\zeta_1}
- \sum_{\zeta_1'\zeta_2'} \sum_{\zeta_3'\zeta_4'} 
\Lambda_{\zeta_2',\zeta_4; \zeta_3,\zeta_1'}(Q) \chi_{\zeta_3',\zeta_2'; \zeta_1',\zeta_4'}(Q) 
\Gamma_{\zeta_2,\zeta_4'; \zeta_3',\zeta_1}^{(0)}, \label{Eq:Lambda}\\ 
\Gamma_{\zeta_2,\zeta_4; \zeta_3,\zeta_1}(Q) = \Gamma_{\zeta_2,\zeta_4; \zeta_3,\zeta_1}^{(0)} 
- \sum_{\zeta_1'\zeta_2'} \sum_{\zeta_3'\zeta_4'} 
\Gamma_{\zeta_2',\zeta_4; \zeta_3,\zeta_1'}(Q) \chi_{\zeta_3',\zeta_2'; \zeta_1',\zeta_4'}(Q) 
\Gamma_{\zeta_2,\zeta_4'; \zeta_3',\zeta_1}^{(0)}, \label{Eq:Gamma}
\end{eqnarray}
where $\Gamma_{\zeta_1,\zeta_2; \zeta_3,\zeta_4}^{(0)}$ is 
the antisymmetrized bare Coulomb interaction given by 
$\Gamma_{\zeta_1,\zeta_2;\zeta_3,\zeta_4}^{(0)} = 
I_{\ell_1,\ell_2;\ell_3,\ell_4}\delta_{\sigma_1\sigma_4}\delta_{\sigma_2\sigma_3} 
- I_{\ell_1,\ell_2;\ell_4,\ell_3}\delta_{\sigma_1\sigma_3}\delta_{\sigma_2\sigma_4}$.  
$\chi(Q)$ is the polarization function calculated by 
\begin{eqnarray}
\chi_{\zeta_3,\zeta_2; \zeta_1,\zeta_4}(Q) &=& \frac{1}{N} \sum_{\mib{k}} \sum_{a,a'} 
u_{\zeta_1, a}(\mib{k}) u_{\zeta_4, a}^*(\mib{k})
u_{\zeta_3, a'}(\mib{k}+\mib{Q}) u_{\zeta_2, a'}^*(\mib{k}+\mib{Q}) \chi_{a,a'}(\mib{k}; Q), \nonumber \\ \\
\chi_{a,a'}(\mib{k}; Q) &=& \frac{n_{a'}(\mib{k}+\mib{Q}) - n_a(\mib{k})}
{\Omega + E_a(\mib{k}) - E_{a'}(\mib{k} + \mib{Q}) + i \epsilon}, 
\end{eqnarray}
where $\epsilon$ is interpreted as the damping rate 
of the excited electron-hole pair near the Fermi level, as already introduced above.
Solving eqs.~(\ref{Eq:Lambda}) and (\ref{Eq:Gamma}), we can determine 
$\Lambda_{\zeta_2,\zeta_4; \zeta_3,\zeta_1}(Q)$ 
and $\Gamma_{\zeta_2,\zeta_4; \zeta_3,\zeta_1}(Q)$ within RPA. 

To resolve into the contribution from each process of the 0th-order, $p$-scattering 
and $d$-scattering, we introduce the process-resolved spectra as follows: 
\begin{eqnarray}
W^{(0)}(q, \mib{e}; q', \mib{e}') &=& \frac{2 \pi}{N} \sum_{\mib{k}_1} \sum_{a_1a_2} 
n_{a_1}(\mib{k}_1) [1-n_{a_2}(\mib{k}_1 + \mib{Q}) ]
\delta (\Omega + E_{a_1}(\mib{k}_1)- E_{a_2}(\mib{k}_1 + \mib{Q})) \nonumber \\
&& \times \biggl| \sum_i^{\rm t.m.u.}
\sum_{\zeta_1\zeta_2}^{@\mib{r}_i} \sum_{JJ'} \sum_{M=-J}^{J} \sum_{M'=-J'}^{J'} 
w_{\zeta_2, JM}(\mib{r}_i; \mib{q}, \mib{e}) 
w_{\zeta_1, J'M'}^*(\mib{r}_i; \mib{q}', \mib{e}') \nonumber \\
&& \times F_{\zeta_2 JM, \zeta_1 J'M'; a_1, a_2}^{(0)}(\mib{r}_i; \mib{k}_1; q, q') 
\biggr|^2, \label{Eq:W0} \\
W^{(p)}(q, \mib{e}; q', \mib{e}') &=& \frac{2 \pi}{N} \sum_{\mib{k}_1} \sum_{a_1a_2} 
n_{a_1}(\mib{k}_1) [1-n_{a_2}(\mib{k}_1 + \mib{Q}) ]
\delta (\Omega + E_{a_1}(\mib{k}_1)- E_{a_2}(\mib{k}_1 + \mib{Q})) \nonumber \\
&& \times \biggl| \sum_i^{\rm t.m.u.}
\sum_{\zeta\zeta'}^{@\mib{r}_i} \sum_{JJ'} \sum_{M=-J}^{J} \sum_{M'=-J'}^{J'} 
w_{\zeta, JM}(\mib{r}_i; \mib{q}, \mib{e}) w_{\zeta', J'M'}^*(\mib{r}_i; \mib{q}', \mib{e}') \nonumber\\
&& \times \sum_{\zeta_1\zeta_2} u_{\zeta_2, a_2}^*(\mib{k}_1+\mib{Q})u_{\zeta_1, a_1}(\mib{k}_1) 
F_{\zeta JM, \zeta' J'M'; \zeta_1, \zeta_2}^{(p)}(\mib{r}_i; q, q') \biggr|^2, \label{Eq:Wp}\\ 
W^{(d)}(q, \mib{e}; q', \mib{e}') &=& \frac{2 \pi}{N} \sum_{\mib{k}_1} \sum_{a_1a_2} 
n_{a_1}(\mib{k}_1) [1-n_{a_2}(\mib{k}_1 + \mib{Q}) ]
\delta (\Omega + E_{a_1}(\mib{k}_1)- E_{a_2}(\mib{k}_1 + \mib{Q})) \nonumber \\
&& \times \biggl| \sum_i^{\rm t.m.u.} 
\sum_{\zeta\zeta'}^{@\mib{r}_i} \sum_{JJ'} \sum_{M=-J}^{J} \sum_{M'=-J'}^{J'} 
w_{\zeta, JM}(\mib{r}_i; \mib{q}, \mib{e}) w_{\zeta', J'M'}^*(\mib{r}_i; \mib{q}', \mib{e}') \nonumber\\
&& \times \sum_{\zeta_1\zeta_2} u_{\zeta_2, a_2}^*(\mib{k}_1+\mib{Q})u_{\zeta_1, a_1}(\mib{k}_1) 
F_{\zeta JM, \zeta' J'M'; \zeta_1, \zeta_2}^{(d)}(\mib{r}_i; q, q') \biggr|^2, \label{Eq:Wd}
\end{eqnarray}
These are obtained from eq.~(\ref{Eq:W2}) by keeping only one of 
$F^{(0)}_{\zeta JM, \zeta' J'M'; a_1, a_2}(\mib{r}_i; \mib{k}_1; q, q')$, 
$F^{(p)}_{\zeta JM, \zeta' J'M'; \zeta_1, \zeta_2}(\mib{r}_i; q, q')$ and 
$F^{(d)}_{\zeta JM, \zeta' J'M'; \zeta_1, \zeta_2}(\mib{r}_i; q, q')$ and 
setting the rest two to zero in eq.~(\ref{Eq:F}). 

Further to resolve orbital-excitation processes involved in the 0th-order, 
$p$-scattering and $d$-scattering processes, 
we introduce the orbital-spin-resolved spectra as follows: 
\begin{eqnarray}
W_{\zeta_1 \rightarrow \zeta_2}^{(0)}(q, \mib{e}; q', \mib{e}') 
&=& \frac{2 \pi}{N} \sum_{\mib{k}_1} \sum_{a_1a_2} 
n_{a_1}(\mib{k}_1) [1-n_{a_2}(\mib{k}_1 + \mib{Q}) ]
\delta (\Omega + E_{a_1}(\mib{k}_1)- E_{a_2}(\mib{k}_1 + \mib{Q})) \nonumber \\
&& \times \biggl| \sum_{JJ'} \sum_{M=-J}^{J} \sum_{M'=-J'}^{J'}  
w_{\zeta_2, JM}(\mib{r}_i; \mib{q}, \mib{e}) w_{\zeta_1, J'M'}^*(\mib{r}_i; \mib{q}', \mib{e}') \nonumber \\
&& \times F_{\zeta_2 JM, \zeta_1 J'M'; a_1, a_2}^{(0)}(\mib{r}_i; \mib{k}_1; q, q') \biggr|^2. 
\label{Eq:W0ll} \\
W_{\zeta_1 \rightarrow \zeta_2}^{(p)}(q, \mib{e}; q', \mib{e}') 
&=& \frac{2 \pi}{N} \sum_{\mib{k}_1} \sum_{a_1a_2} 
n_{a_1}(\mib{k}_1) [1-n_{a_2}(\mib{k}_1 + \mib{Q}) ]
\delta (\Omega + E_{a_1}(\mib{k}_1)- E_{a_2}(\mib{k}_1 + \mib{Q})) \nonumber \\
&& \times \biggl| \sum_i^{\rm t.m.u.} 
\sum_{\zeta\zeta'}^{@\mib{r}_i} \sum_{JJ'} \sum_{M=-J}^{J} \sum_{M'=-J'}^{J'} 
w_{\zeta, JM}(\mib{r}_i; \mib{q}, \mib{e}) w_{\zeta', J'M'}^*(\mib{r}_i; \mib{q}', \mib{e}') \nonumber\\
&& \times u_{\zeta_2, a_2}^*(\mib{k}_1+\mib{Q})u_{\zeta_1, a_1}(\mib{k}_1) 
F_{\zeta JM, \zeta' J'M'; \zeta_1, \zeta_2}^{(p)}(\mib{r}_i; q, q') \biggr|^2, \label{Eq:Wpll}\\ 
W_{\zeta_1 \rightarrow \zeta_2}^{(d)}(q, \mib{e}; q', \mib{e}') 
&=& \frac{2 \pi}{N} \sum_{\mib{k}_1} \sum_{a_1a_2} 
n_{a_1}(\mib{k}_1) [1-n_{a_2}(\mib{k}_1 + \mib{Q}) ]
\delta (\Omega + E_{a_1}(\mib{k}_1)- E_{a_2}(\mib{k}_1 + \mib{Q})) \nonumber \\
&& \times \biggl| \sum_i^{\rm t.m.u.} 
\sum_{\zeta\zeta'}^{@\mib{r}_i} \sum_{JJ'} \sum_{M=-J}^{J} \sum_{M'=-J'}^{J'} 
w_{\zeta, JM}(\mib{r}_i; \mib{q}, \mib{e}) w_{\zeta', J'M'}^*(\mib{r}_i; \mib{q}', \mib{e}') \nonumber\\
&& \times u_{\zeta_2, a_2}^*(\mib{k}_1+\mib{Q})u_{\zeta_1, a_1}(\mib{k}_1) 
F_{\zeta JM, \zeta' J'M'; \zeta_1, \zeta_2}^{(d)}(\mib{r}_i; q, q') \biggr|^2, \label{Eq:Wdll}
\end{eqnarray}
where, in every of eqs.~(\ref{Eq:W0ll}), ~(\ref{Eq:Wpll}) and ~(\ref{Eq:Wdll}), 
the orbital-spin index $\zeta_1$ [$\zeta_2$] represents the orbital and spin 
of the electron removed below $E_F$ [left above $E_F$] in the final state 
at $t=+\infty$, where $E_F$ is the Fermi energy and $t$ is time. 
Throughout the study, we specify spin states by taking the spin axis parallel to the $z$-axis. 
Equations, (\ref{Eq:W0ll}), (\ref{Eq:Wpll}) and (\ref{Eq:Wdll}) are obtained 
by suspending the summation with respect to orbital-spin indices $\zeta_1$ and $\zeta_2$ 
in the right-hand side of eqs.~(\ref{Eq:W0}), ~(\ref{Eq:Wp}) and ~(\ref{Eq:Wd}). 
In eq.~(\ref{Eq:W0ll}), if we select orbitals $\zeta_1$ and $\zeta_2$ specifically, then atomic site $\mib{r}_i$ 
in the unit cell is determined uniquely to give non-vanishing contributions, 
since the factor $w_{\zeta_2, JM}(\mib{r}_i; q,e) w_{\zeta_1, J'M'}^*(\mib{r}_i; q',e')$ 
does not vanish only in the case where both of $\zeta_1$ and $\zeta_2$ reside on atomic site $\mib{r}_i$. 
Thus summation in $\mib{r}_i$ does not appear in eq.~(\ref{Eq:W0ll}). 
On the other hand, in eqs.~(\ref{Eq:Wpll}) and ~(\ref{Eq:Wdll}), 
summation with respect to $\mib{r}_i$ is necessary, since $\zeta_1$ and $\zeta_2$ 
need not reside on the site $\mib{r}_i$ where x-ray absorption and emission occur. 
All of the above formulae are valid also for general cases including several transition-metal atoms.

Here we should note that the total RIXS intensity 
$W(q, \mib{e}; q', \mib{e}')$ does not equal the sum 
of the resolved intensities, e.g., $W(q, \mib{e}; q', \mib{e}') 
\neq W^{(0)}(q, \mib{e}; q', \mib{e}')  + W^{(p)}(q, \mib{e}; q', \mib{e}') + W^{(d)}(q, \mib{e}; q', \mib{e}') $, 
$W^{(p)}(q, \mib{e}; q', \mib{e}')  \neq \sum_{\zeta_1\zeta_2} 
W_{\zeta_1 \rightarrow \zeta_2}^{(p)}(q, \mib{e}; q', \mib{e}') $, 
$W^{(d)}(q, \mib{e}; q', \mib{e}')  \neq \sum_{\zeta_1\zeta_2} 
W_{\zeta_1 \rightarrow \zeta_2}^{(d)}(q, \mib{e}; q', \mib{e}') $, and so on. 
This is because the total summed spectrum $W(q, \mib{e}; q', \mib{e})$ contains interference terms 
such as $F^{(0)}(\mib{r}_i; \mib{k}_1; q, q') F^{(d)}{}^*(\mib{r}_i; q, q')$, 
while resolved spectra $W^{(0)}(q, \mib{e}; q', \mib{e}')$ and $W^{(d)}(q, \mib{e}; q', \mib{e}') $ contain 
only $|F^{(0)}(\mib{r}_i; \mib{k}_1; q, q')|^2$ and $|F^{(d)}(\mib{r}_i; q, q')|^2$, respectively. 
Nevertheless these resolved spectra are useful to take microscopic insights 
into the mechanism of RIXS, as we shall see later. 

For numerical calculation of eq.~(\ref{Eq:W2}), 
we use the Lorentzian expression for the $\delta$-function: 
\begin{equation}
\delta (\Omega + E_{a_1}(\mib{k}_1)- E_{a_2}(\mib{k}_1 + \mib{Q})) \rightarrow 
\frac{1}{\pi} \frac{\epsilon} 
{[\Omega + E_{a_1}(\mib{k}_1)- E_{a_2}(\mib{k}_1 + \mib{Q})]^2 + \epsilon^2}. 
\end{equation}
This function possesses poles 
at $\Omega = E_{a_2}(\mib{k}_1 + \mib{Q}) - E_{a_1}(\mib{k}_1) \pm i \epsilon$, 
which correspond to the transition from band $a_1$ to band $a_2$. 
Therefore, at a first glance, one might consider that eq.~(\ref{Eq:W2}) 
describes only simple band-to-band transitions and fails to describe local $d$-$d$ transitions. 
This naive view is not correct, as explained next. 
We should note that, for overall consistency, the factor $\epsilon$ should 
equal the damping rate of the excited electron-hole pair, already introduced above. 
The position of the pole $\Omega = E_{a_2}(\mib{k}_1 + \mib{Q}) - E_{a_1}(\mib{k}_1) \pm i \epsilon$ 
is modified to a non-trivial position by the RPA correction. 
The modified poles describe bound states between the excited electron and hole in the final state. 
As a result, we shall see not only charge-transfer excitations but also magnon excitation 
and local $d$-$d$ excitations can be described within our HF-RPA calculation 
on the basis of electronic bands. 

For later discussions, we define the following quantity, 
\begin{eqnarray}
P_{\zeta \zeta'}^J(\mib{q}, \mib{e}; \mib{q}', \mib{e}')  
& \equiv & \biggl| \sum_{M=-J}^{J} w_{\zeta, JM}(\mib{r}_i; \mib{q}, \mib{e}) 
w_{\zeta', JM}^*(\mib{r}_i; \mib{q}', \mib{e}') \biggr|^2 \nonumber \\
& \propto & \biggl| \sum_{M=-J}^{J} 
\langle d_\zeta | \mib{e} \cdot \mib{r}| 2p_{JM} \rangle 
\langle 2p_{JM} | \mib{e}' \cdot \mib{r}| d_{\zeta'} \rangle \biggr|^2,  
\label{Eq:P} 
\end{eqnarray}
where $\zeta = (\ell,\sigma)$, $\zeta' = (\ell', \sigma')$, 
and $\ell$ and $\ell'$ lie on transition-metal site $\mib{r}_i$. 
We should note that $\zeta$ [$\zeta'$] represents the orbital and spin of the $d$ electron 
which is initially promoted from a $2p$ state [finally decays down to the empty $2p$ state], 
and this electron could be different in general from the electron 
left above $E_F$ [removed below $E_F$] in the final state at $t = + \infty$. 
Nevertheless, we shall see $P_{\zeta \zeta'}^J(\mib{q}, \mib{e}; \mib{q}', \mib{e}') $ 
is useful for understanding the dependence of RIXS spectra on the scattering geometry. 

\section{Numerical Results}
\label{Sc:Results}

\subsection{Overall spectral structure}
\label{Sc:overall}

\begin{figure}
\begin{center}
\includegraphics[width=90mm]{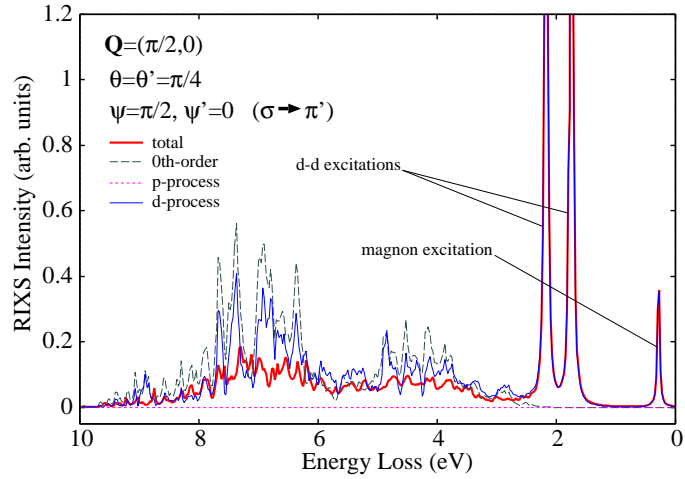}
\end{center}
\caption{
(Color online) 
Typical calculated spectra as a function of energy loss. 
In-plane momentum transfer of photons is set to 
$\mib{Q} = \mib{q}-\mib{q}' = (\pi/2, 0)$. 
Bragg angles are $\theta=\theta'=\pi/4$ (specular geometry). 
Initial incoming photons are in $\sigma$ polarization (i.e., $\psi = \pi/2$)  
and the final outgoing photons are in $\pi$ polarization (i.e., $\psi' = 0$). 
The thick solid curve is the total spectra. 
The thin dashed, dotted and solid lines are the process-resolved spectra, 
$W^{(0)}(q, \mib{e}; q', \mib{e}')$ (0th-order process), 
$W^{(p)}(q, \mib{e}; q', \mib{e}')$ ($p$-process), 
and $W^{(d)}(q, \mib{e}; q', \mib{e}')$ ($d$-process), respectively. 
$W^{(p)}(q, \mib{e}; q', \mib{e}')$ is negligibly small. 
The thick solid and thin solid curves almost completely merge 
at the low-energy region below 2.2 eV. }
\label{Fig:rixsspectraovrall}
\end{figure}

Throughout our present study, we set the incident photon energy 
at the Cu-$L_3$ edge (see Appendix B for absorption spectra), 
and set the azimuth angles as $\phi=\phi'=0$. 
Firstly, we present a typical calculated result of RIXS spectrum 
within our theoretical framework (see the thick solid curve in Fig. ~\ref{Fig:rixsspectraovrall}). 
The spectrum consists of three main features: magnon excitation 
at low energies up to 400 meV, $d$-$d$ orbital excitations around 1.7-2.1 eV, 
and the charge-transfer excitations between 3-9 eV, 
which are semi-quantitatively consistent with experimental spectra~\cite{Sala2011}. 

The origin of each feature is clarified by using the process-resolved spectra 
(see \S~\ref{Sc:RIXSintensity}, and eqs.~(\ref{Eq:W0}), (\ref{Eq:Wp}) and (\ref{Eq:Wd})). 
The low-energy $d$-$d$ and magnon excitation weights originate solely from the $d$-process. 
We should note that, in our electronic band approach, 
the $d$-$d$ excitations are described as bound states of electron and hole on the bands. 
Therefore $d$-$d$ excitations are not contributed from the 0th-order process 
but from the $d$-process. 
The 0th-order process contributes only above 2.2 eV excitation energy to the RIXS weight, 
since the insulating gap is 2.2 eV. 
On the other hand, high-energy (above 3 eV) part is contributed 
from both the 0th-order process and $d$-process. 
The $p$-process, in which the $d$ electrons are excited 
by screening the inner-shell $2p$ core hole, is negligible in the present case. 
As intuitively understood by the local picture, the Cu-$d$ shell 
is completely filled with 10 electrons in the intermediate state, 
and the core-hole cannot be screened by the Cu-$d$ electrons. 
In this point, $L$-edge RIXS substantially differs from $K$-edge RIXS. 
In $K$-edge RIXS, screening of the inner-shell $1s$ core hole 
is the essential process~\cite{Nomura2005}. 
Thus the Slater-Condon integrals between the $2p$ and $3d$ states, 
$F^{0}(p,d)$ and $F^{2}(p,d)$, are not significant parameters in the present Cu  $L$-edge case. 
For the same reason, $G^{1}(p,d)$ and $G^{3}(p,d)$ are not significant.

\subsection{Magnon excitation}
\label{Sc:magnon}

\begin{figure}
\begin{center}
\includegraphics[width=90mm]{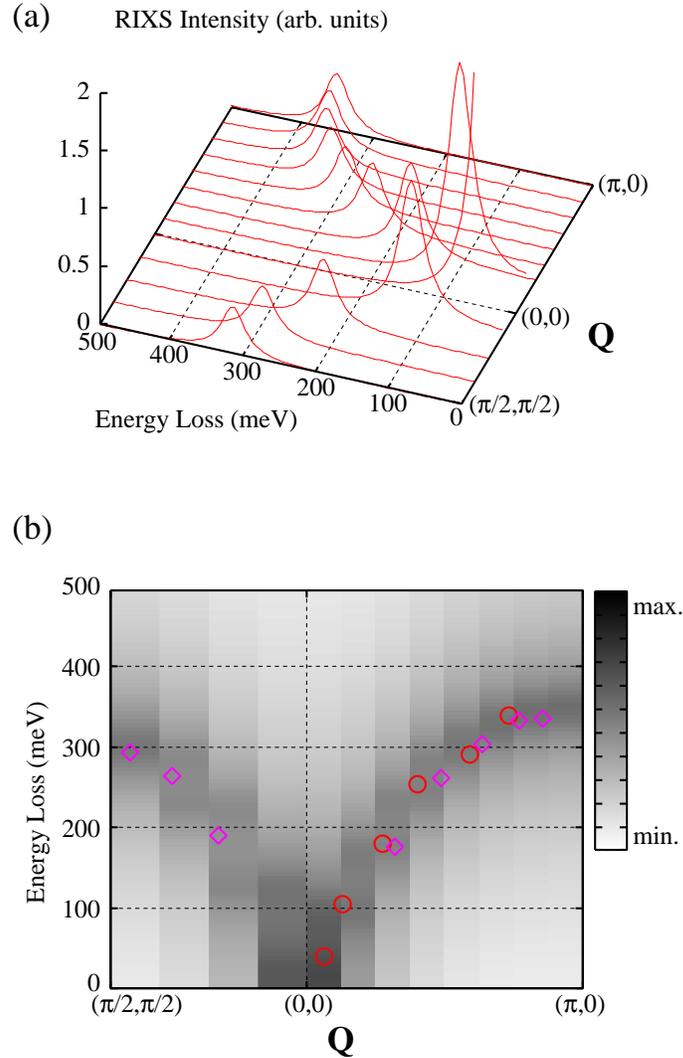}
\end{center}
\caption{
(Color online)
(a) Low-energy part of calculated RIXS spectra is depicted 
along the symmetry lines in the Brillouin zone. 
(b) Gray-level map of the low-energy RIXS intensity. 
The scattering geometry for calculation is 
$\theta = \theta' = \pi/4$ and $\pi \rightarrow \sigma'$ 
($\psi = 0$, $\psi'=\pi/2$). 
The plots are experimental data of magnon excitation energy, 
which are read from Refs.~\ref{Ref:Braicovich2010a} and \ref{Ref:Dean2012}. 
Experimental data contain those of $\theta +\theta' = 90^\circ$ 
(for small $|\mib{Q}|$) and $130^\circ$ (for large $|\mib{Q}|$).} 
\label{Fig:magnon}
\end{figure}

Here we focus on the low-energy feature below 0.5 eV. 
As we have mentioned, magnon (spin-wave) mode appears 
below about 350 meV, and is dispersive as shown in Fig.~\ref{Fig:magnon}(a). 
The theoretical dispersion of the magnon is quite consistent 
with experimental one in quantitative level as shown 
in Fig.~\ref{Fig:magnon}(b). 
This magnon mode originates from spin-flip processes in the $d_{x^2-y^2}$ orbital. 
Here note that we have started not from a Heisenberg spin model 
but from an electron band model to derive the magnon spectrum. 
To fit well to the experimental magnon dispersion, Heisenberg-model approaches  
require not only nearest-neighbor exchange $J$ but also long-range ones. 
The exchange parameters are often used as fitting parameters. 
On the other hand, in our band approach, effects of the long-range 
magnetic correlations are taken into account 
through the inter-site hopping parameters, which are determined 
by the first-principles band calculation. 
It is notable that the magnon intensity should increase divergently towards $\mib{Q}=0$. 
This divergence is not a result from ordinary elastic processes, 
since we have already excluded the elastic contributions, 
which are expressed by diagrams whose upper normally-time-ordered and lower 
reversely-time-ordered branches are disconnected (See Fig.~\ref{Fig:diagrams}(I)). 

\begin{figure}
\begin{center}
\includegraphics[width=90mm]{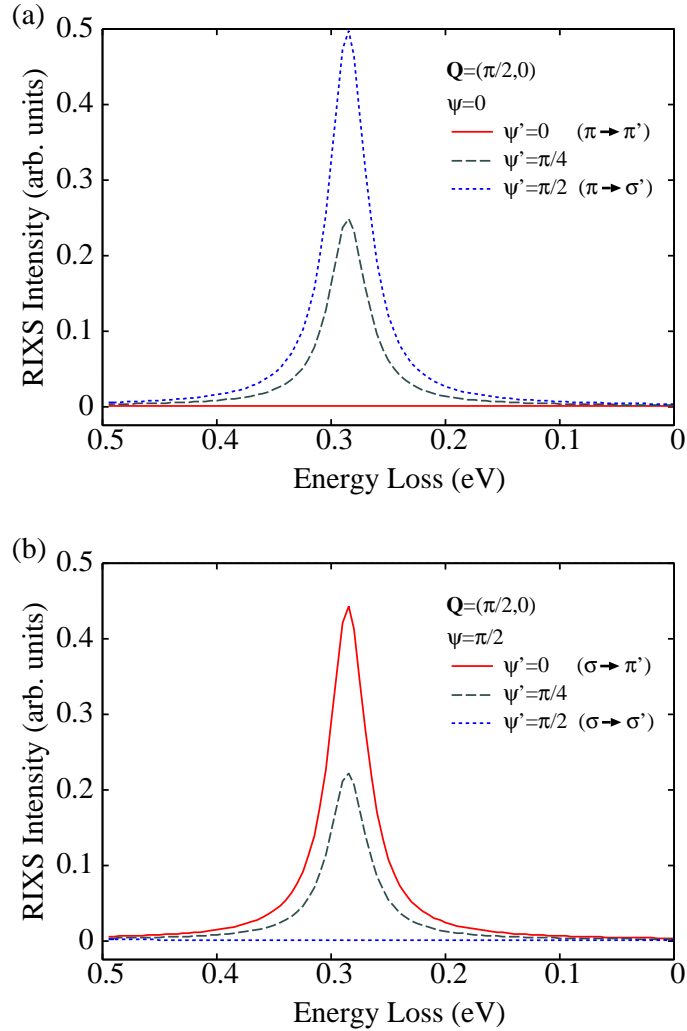}
\end{center}
\caption{
(Color online) 
Polarization dependence of the low-energy part of calculated RIXS spectrum. 
$\theta = \theta' = \pi/4$, and in-plane momentum transfer of photons is set 
to $\mib{Q} = \mib{q}-\mib{q}' = (\pi/2, 0)$. 
Initial incoming photons are in $\pi$ polarization (i.e., $\psi = 0$) in (a) 
and in $\sigma$ polarization (i.e., $\psi = \pi/2$) in (b).}
\label{Fig:magnonpoldep}
\end{figure}

Photon-polarization dependence of the magnon excitation is so drastic, 
as shown in Fig.~\ref{Fig:magnonpoldep}. 
Intensity of the magnon excitation is maximized, 
when polarization directions of incoming and outgoing photons 
are perpendicular to each other 
($\pi \rightarrow \sigma'$ or $\sigma \rightarrow \pi'$). 
On the other hand, the weight of the magnon excitation vanishes, 
when polarization directions of incoming and outgoing photons 
are identical to each other ($\pi \rightarrow \pi'$ or $\sigma \rightarrow \sigma'$). 
Thus, it may become possible to identify the magnon weight more clearly, 
by sorting out not only incoming photons but also outgoing photons 
with respect to polarization. 

\subsection{$d$-$d$ excitations}
\label{Sc:dd}

\begin{figure}
\begin{center}
\includegraphics[width=100mm]{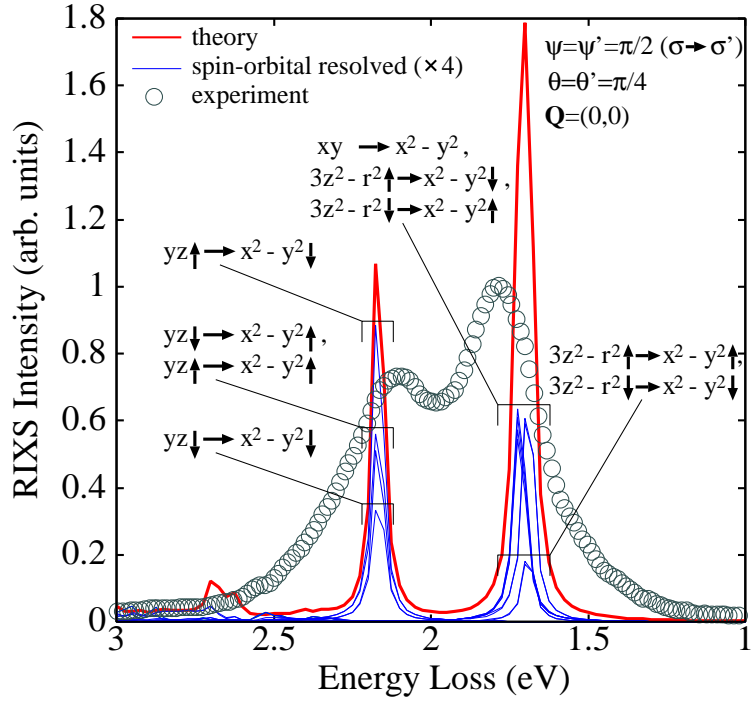}
\end{center}
\caption{(Color online)
A typical $d$-$d$ excitation spectrum around 2 eV. 
Thick and thin solid curves denote the calculated results 
of total and spin-orbital-resolved RIXS spectra, respectively. 
For the calculated results, polarization of photons is set as 
$\psi = \psi' = \pi/2$ ($\sigma \rightarrow \sigma'$). 
For the weight from $xy \rightarrow x^2-y^2$, four curves 
($xy \uparrow \rightarrow x^2-y^2 \uparrow, 
xy \uparrow \rightarrow x^2-y^2 \downarrow, 
xy \downarrow \rightarrow x^2-y^2 \uparrow, 
xy \downarrow \rightarrow x^2-y^2 \downarrow$) almost evenly 
contribute and their curves are overlaid (Note that we are working 
with the electron representation not with the hole one). 
Circles are experimental data read from Ref.~\ref{Ref:Sala2011} 
(incident incoming photons are in $\sigma$ polarization, 
but polarization of final outgoing photons is not discriminated in the experiment). }
\label{Fig:ddresolved}
\end{figure}

Here we focus on the $d$-$d$ excitations around 2 eV energy loss. 
In Fig.~\ref{Fig:ddresolved}, a calculated spectrum for a typical scattering geometry 
is compared with the experimental data read from Ref.~\ref{Ref:Sala2011}. 
Two-peak structure is well reproduced by the present calculation, 
although the peak positions depend on the modification 
of the completely-filled Cu-$d$ levels (see Appendix A). 
In that scattering geometry, the low-energy peak around 1.7 eV are attributed mainly 
to the $d$-$d$ transition from $d_{xy}$ to $d_{x^2-y^2}$, while the high-energy peak 
around 2.2 eV is to the $d$-$d$ transition from $d_{yz}$ to $d_{x^2-y^2}$ 
(Note that we are working with the electron representation not with the hole one). 
The transition from $d_{xz}$ to $d_{x^2-y^2}$ is absent in that scattering geometry. 
The interval between the two peaks is somewhat overestimated. 
We consider that this overestimation is ascribed to the band structure calculation 
(and subsequent Wannier fitting and Hartree-Fock calculation), 
since we retained the energy level splitting between the $d_{xy}$ and $d_{yz/xz}$ levels 
in modifying the completely filled Cu-$d$ levels (see Appendix A). 
In Fig.~\ref{Fig:ddresolved}, we should note that the peaks are not contributed 
only from spin-conserved excitations but also from spin-flipped excitations. 
In particular, the spin-conserved and spin-flipped transitions 
from $d_{xy}$ to $d_{x^2-y^2}$ almost evenly occur. 
Energy positions of $d$-$d$ excitation peaks do not significantly depend 
on momentum transfer, reflecting their localized nature 
(Momentum-transfer dependence is not shown). 

\begin{figure}
\begin{center}
\includegraphics[width=90mm]{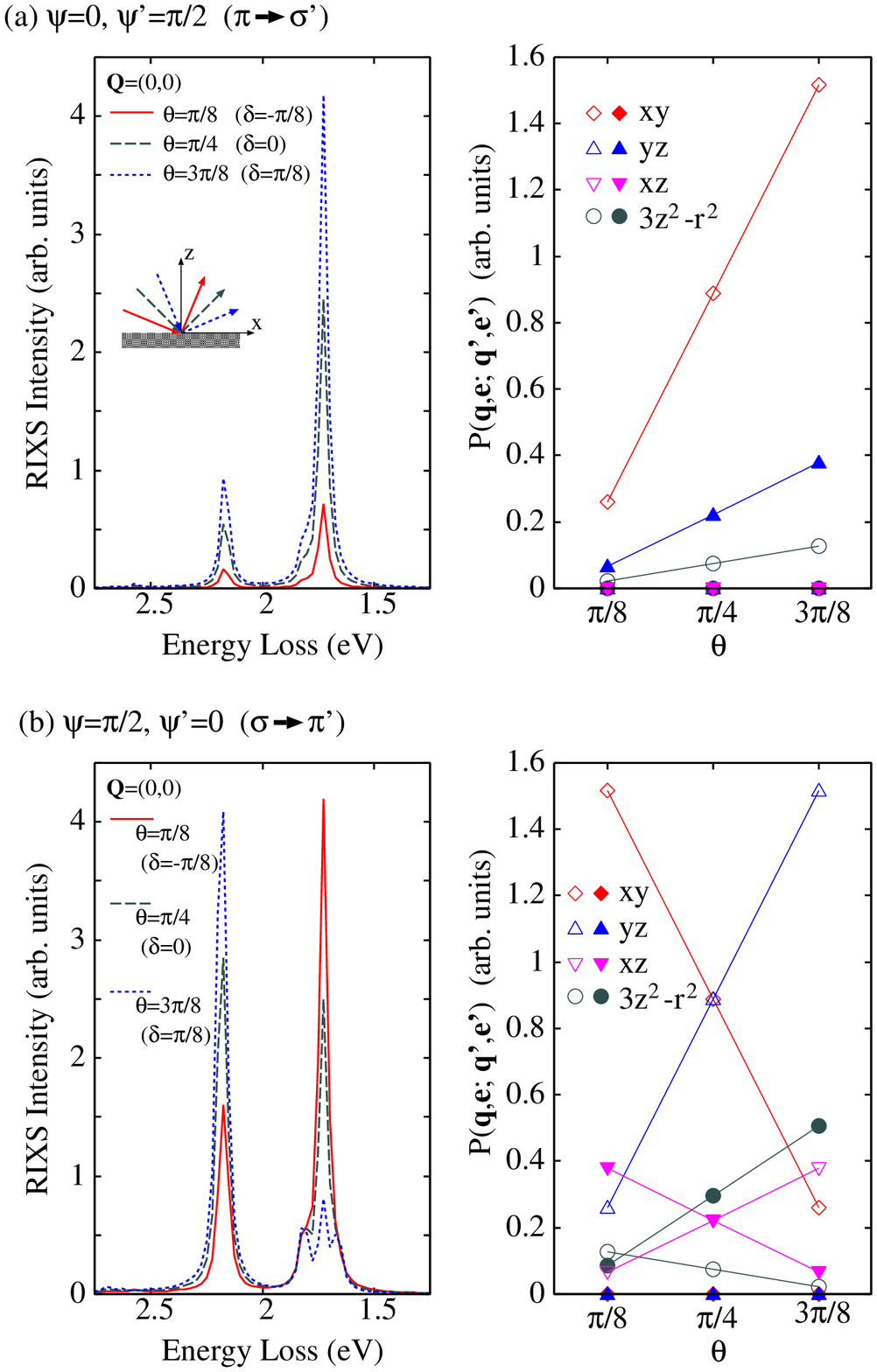}
\end{center}
\caption{
(Color online)
Dependence of the $d$-$d$ excitation spectrum (left panels) 
and $P_{(d_{x^2-y^2} \, \sigma)(\ell'\sigma')}^{J=3/2}(\mib{q}, \mib{e}; \mib{q}', \mib{e}')$ 
(right panels) on the Bragg angle $\theta$ (or dependence on $\delta = (\theta - \theta')/2$) 
with maintaining $\theta + \theta' = \pi/2$. 
$\theta = \pi/8$, $\pi/4$, $3\pi/8$ correspond to $\delta = -\pi/8, 0, +\pi/8$, respectively. 
In the right panels, 
$P_{(x^2-y^2 \, \sigma)(\ell'\sigma')}^{J=3/2}(\mib{q}, \mib{e}; \mib{q}', \mib{e}')$ 
is shown for $\ell' = xy, yz, xz, 3z^2-r^2$. 
Empty and filled symbols are for the spin-conserved ($\sigma = \sigma'$) 
and spin-flipped ($\sigma = -\sigma'$) components, respectively. 
In (a), photon polarizations are set as $\psi = 0$ and $\psi'=\pi/2$ ($\pi \rightarrow \sigma'$). 
In (b), photon polarizations are set as $\psi = \pi/2$ and $\psi'=0$ ($\sigma \rightarrow \pi'$). 
Inset in (a) schematically represents the scattering geometry for the three incident Bragg angles, 
$\theta = \pi/8, \pi/4, 3\pi/8 $. }
\label{Fig:ddthetadep}
\end{figure}

\begin{figure}
\begin{center}
\includegraphics[width=90mm]{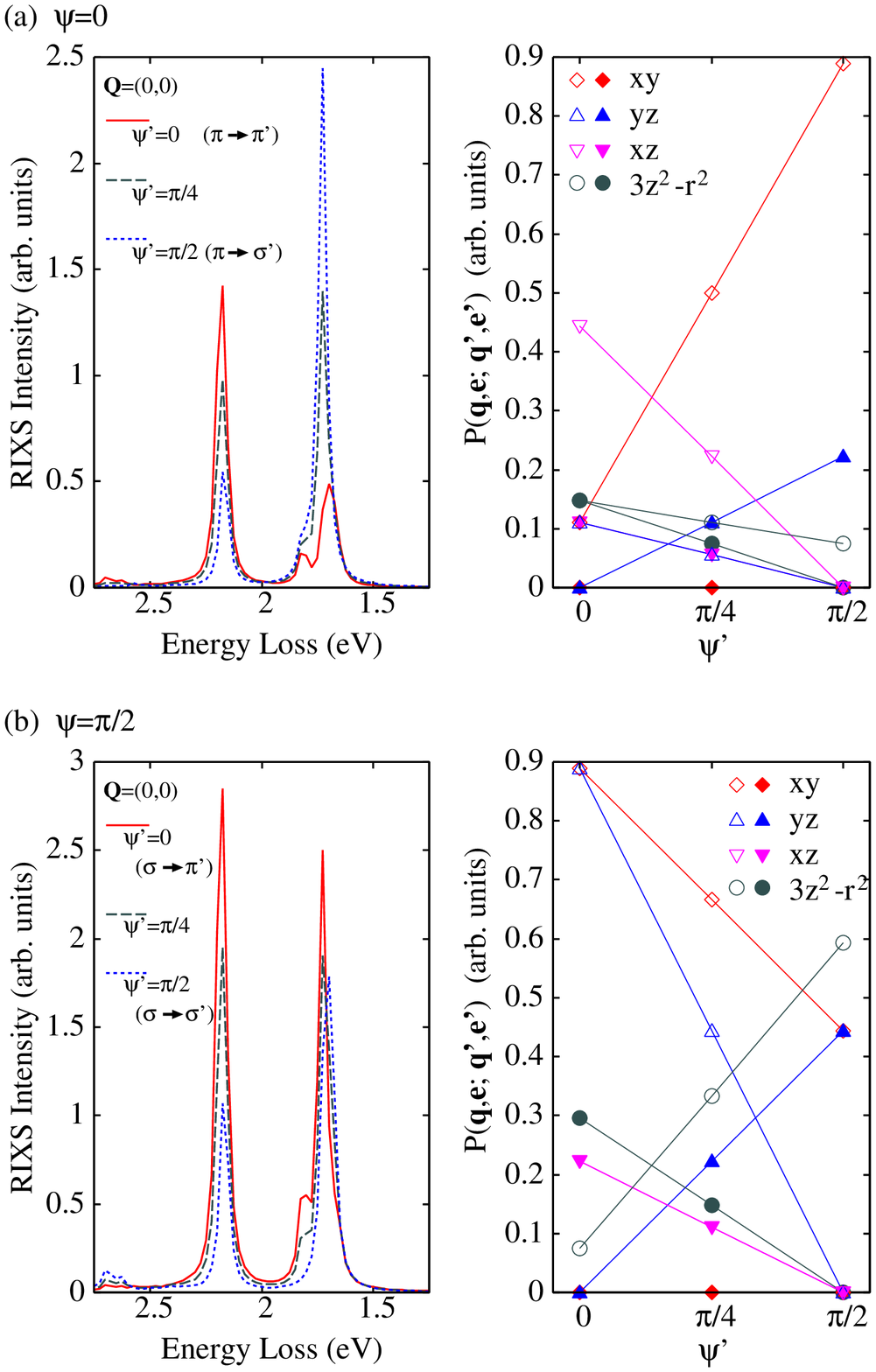}
\end{center}
\caption{
(Color online)
Dependence of the $d$-$d$ excitation spectrum (left panels) and 
$P_{(x^2-y^2 \, \sigma)(\ell'\sigma')}^{J=3/2}(\mib{q}, \mib{e}; \mib{q}', \mib{e}')$ 
(right panels) on polarizations of incoming and outgoing photons. In the right panels, 
$P_{(x^2-y^2 \, \sigma)(\ell'\sigma')}^{J=3/2}(\mib{q}, \mib{e}; \mib{q}', \mib{e}')$ 
is shown for $\ell' = xy, yz, xz, 3z^2-r^2$. 
Empty and filled symbols are for the spin-conserved ($\sigma = \sigma'$) 
and spin-flipped ($\sigma = -\sigma'$) components, respectively. 
In (a) [in (b)], the incident incoming photons are in $\pi$ [$\sigma$] polarization, i.e., $\psi=0$ [$\psi=\pi/2$]. 
The Bragg angles are $\theta = \theta' = \pi/4$ in the both cases.}
\label{Fig:ddpoldep}
\end{figure}

The $d$-$d$ excitation spectrum significantly depends on the scattering geometry. 
In the right panels of Fig.~\ref{Fig:ddthetadep}, the $d$-$d$ excitation spectra 
are displayed for various Bragg angles with $\theta + \theta' = \pi/2$. 
For $\pi \rightarrow \sigma'$, $d$-$d$ excitation intensity monotonically 
increases with increasing $\theta$ (left panel of Fig.~\ref{Fig:ddthetadep}(a)). 
For $\sigma \rightarrow \pi'$, intensity of the 1.7 eV [2.2 eV] peak 
decreases [increases] drastically with increasing $\theta$ 
(left panel of Fig.~\ref{Fig:ddthetadep}(b)). 
These behaviors seem well correlated with 
$P_{(x^2-y^2 \, \sigma)(\ell'\sigma')}^{J=3/2}(\mib{q}, \mib{e}; \mib{q}', \mib{e}')$: 
As seen in Fig.~\ref{Fig:ddresolved}, the 1.7 eV peak originates mainly 
from the $d$-$d$ transition $xy \rightarrow x^2-y^2$, 
and the 2.2 eV peak originates from the $d$-$d$ transition $yz \rightarrow x^2-y^2$. 
For $\pi \rightarrow \sigma'$, $P_{(x^2-y^2 \, \sigma)(xy \, \sigma)}^{J=3/2}$ 
and $P_{(x^2-y^2 \, \sigma)(yz \, -\sigma)}^{J=3/2}$ both increase 
with increasing $\theta$ (right panel of Fig.~\ref{Fig:ddthetadep}(a)). 
For $\sigma \rightarrow \pi'$, $P_{(x^2-y^2 \, \sigma)(xy \, \sigma)}^{J=3/2}$ 
decreases and $P_{(x^2-y^2 \, \sigma)(yz \, \sigma)}^{J=3/2}$ increases 
with increasing $\theta$ (right panel of Fig.~\ref{Fig:ddthetadep}(b)). 
In the left panels of Fig.~\ref{Fig:ddpoldep}, the $d$-$d$ excitation spectra 
are displayed for various polarizations of the incoming and outgoing photons. 
The dependence on polarization can roughly be understood 
in the same manner as above. 
For example, when the incident photon is in $\pi$ polarization ($\psi=0$), 
intensity of the 1.7 eV [2.2 eV] peak increases [decreases] drastically 
with increasing $\psi'$ (left panel of Fig.~\ref{Fig:ddpoldep}(a)). 
This behavior is understood well by noticing 
that $P_{(x^2-y^2 \, \sigma)(xy \, \sigma)}^{J=3/2}$ increases 
and $P_{(x^2-y^2 \, \sigma)(yz \, \sigma)}^{J=3/2}$ decreases 
with increasing $\psi'$ (right panel of Fig.~\ref{Fig:ddpoldep}(a)). 

\section{Discussions}
\label{Sc:Discussions}

In this section, we present some remarks on $L$-edge RIXS, our formulation and calculated results. 

As we have seen, in $L$-edge RIXS, the $d$-$d$ excitations play a major role in the RIXS weight. 
This is in strong contrast to the case of $K$-edge RIXS. 
In the case of $K$-edge RIXS, the process of screening the created $1s$ hole 
by the $d$ electrons is dominant, where the $d$ electrons are scattered isotropically 
due to the strongly localized $1s$ core hole and usually prohibited to change their orbital state. 
In those processes, the $d$ electrons conserve their orbital angular momentum. 
This means that the orbital-diagonal $d$-$d$ excitations are dominant in $K$-edge RIXS. 
As a result, completely filled $d$ orbitals in the initial state are completely filled still 
in the final state of RIXS, and completely empty $d$ orbitals are empty still in the final state. 
Therefore we can construct a good effective theory by neglecting completely filled 
or empty $d$ orbitals, when we describe $K$-edge RIXS. 
In fact, orbital excitations between different $d$ orbitals seem inactive 
in most cases of $K$-edge RIXS, and we can take only active $d$ orbitals, 
e.g., the $d_{x^2-y^2}$ orbital in the case of cuprates with the tetragonal or octahedral coordination. 
On the other hand, in $L$-edge RIXS, orbital states of $d$ electrons can change in the final state. 
This is because the $d$ electrons can exchange their orbital angular momentum 
into spin angular momentum and vice versa, mediated by the $2p$ states in which spin 
and orbital angular momenta are coupled. 
Therefore, to analyze $L$-edge RIXS spectra over a wide-energy range, 
we should include $d$ orbitals which are initially completely filled or completely empty, 
in addition to partially filled $d$ orbitals. 
In this sense, $L$-edge RIXS is essentially a multi-orbital phenomenon, 
and can involve much more complex excitation processes than $K$-edge RIXS. 

The $p$-process, where the $2p$ core hole is screened 
by the Cu-$d$ electrons, is almost ineffective 
in the present case of Cu $L$-edge RIXS. 
This is because the narrow Cu-$d$ band is completely filled 
in the intermediate state of RIXS, and is in great contrast to the case of $K$-edge RIXS. 
However, we should note that the $p$-process can be effective 
in other transition-metal compounds whose transition-metal $d$ bands 
are broad (i.e., localization is weak) and are not completely filled in the intermediate state. 

As we have seen in \S~\ref{Sc:magnon}, 
the magnon intensity becomes divergent towards $\mib{Q}=0$.  
Very recently, Igarashi and Nagao found that such divergence essentially appears 
around $\mib{Q}=0$ as a result from broken symmetry~\cite{Igarashi2015}. 
According to them, this divergence originates from anisotropic terms 
of the scattering amplitude which are a natural result from the long-range ordering, 
and is not reproduced within the so-called fast collision approximation (FCA). 
Although it is beyond the scope of the present work and not straightforward 
to clarify the analytic relation between our present calculation and theirs, 
such anisotropic terms are included within the HF approximation in our present calculation. 
Our calculation suggests that this divergent behavior should survive 
for realistic values of core-hole life time. 
Such divergent behavior around $\mib{Q}=0$ is a substantial difference 
between RIXS and neutron scattering, and deserves to be searched for carefully, 
although the measurement should be performed at low temperatures well 
below the N\'{e}el temperature and the inelastic weight around $\mib{Q}=0$ 
may be difficult to distinguish from the elastic one experimentally. 

Note that the multi-magnon processes are beyond the present theoretical approach. 
$n$-magnon generation processes are represented by diagrams in which the upper 
and lower branches are connected by $2n$ lines, 
which are omitted in Fig.~\ref{Fig:diagrams}(I). 
Inclusion of the multi-magnon excitation processes is not feasible 
for the present complex electronic structure and remains an interesting future work. 

In \S~\ref{Sc:dd}, we have interpreted the dependence of the $d$-$d$ excitation spectrum 
on the scattering geometry by using the square of the product of the dipole-transition matrix, 
$P_{\zeta \zeta'}^J(\mib{q}, \mib{e}; \mib{q}', \mib{e}')$ (see eq.~(\ref{Eq:P})). 
We should note that such interpretation is not always valid, since RIXS intensity 
is not determined only by the dipole-transition matrix elements. 
Analysis based on $P_{\zeta \zeta'}^J(\mib{q}, \mib{e}; \mib{q}', \mib{e}')$ 
will become less effective for systems where orbitals are strongly hybridized 
with each other due to low crystal symmetry. In general, 
the $d$ electron which is initially promoted from a $2p$ state 
can be quite different from the $d$ electron remaining above $E_F$ in the final state, 
as a result from the intermediate scattering processes and orbital hybridization. 

\section{Conclusion}
\label{Sc:Conclusion}

We have discussed the Cu $L_3$-edge RIXS for a typical parent compound 
of high-$T_{\rm c}$ cuprate superconductors La$_2$CuO$_4$ 
on the basis of the first-principles electronic bands. 
We consider that the advantages of our approach are: 
it can reproduce RIXS features in the wide-energy range, 
including the low-energy magnon spectrum, intermediate-energy $d$-$d$ excitations 
and high-energy charge-transfer excitations, consistently with experiments. 
It is applicable to relatively complex three-dimensional multi-orbital systems. 
Finite size effects are absent. 
Therefore it is useful for analyzing experimental data for a wide variety 
of transition-metal compounds. 
Disadvantages are that the ground state needs to be described accurately 
within the Hartree-Fock approximation, 
and multi-magnon weights are not included at all. 

In conclusion, we would like to stress the importance of photon polarization dependence: 
the intensity of magnon excitation and the spectral structure of $d$-$d$ excitations 
depend significantly not only on the polarization direction of incident incoming 
photons but also that of outgoing photons. 
Polarization of scattered photons is not discriminated so far 
in most of experiments~\cite{Braicovich2014}. 
A full knowledge of the both polarizations of incoming and outgoing photons 
could become crucial to correctly assign RIXS spectral weights to magnetic and orbital 
excitation processes. 

\begin{acknowledgment}
The author is grateful to Prof. J. Igarashi, Prof. T. Tohyama, Prof. T. Nagao, 
and Dr. K. Ishii for valuable communications. 
\end{acknowledgment}

\appendix

\section{Hartree-Fock Approximation}
\label{Ap:HF}

\begin{figure}
\begin{center}
\includegraphics[width=90mm]{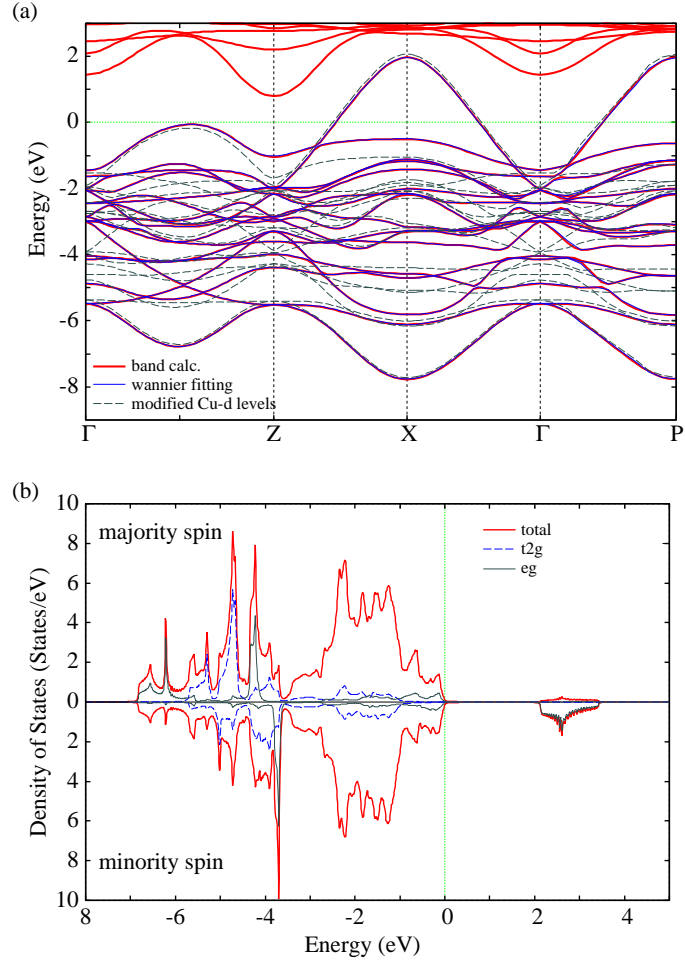}
\end{center}
\caption{
(Color online)
(a) Electronic band structure of the non-magnetic state for La$_2$CuO$_4$. 
Thick solid, thin solid, and dashed curves are the results of the band calculation, 
Wannier-function fitting, and the tight-binding model with modified Cu-$d$ levels 
(see the text for details), respectively. 
The Wannier-fitting curves are almost completely overlaid on the 17 curves 
of the band calculation result around the Fermi energy. 
(b) Density of states (DOS) for the antiferromagnetic ground state. 
Thick solid, thin dashed, and thin solid curves are the results of the total, 
$t_{2g}$ and $e_g$ partial DOS, respectively. The Fermi energy is set to zero 
in the both panels.}
\label{Fig:bandsdos}
\end{figure}

By fitting to the first-principles electronic structure of the nonmagnetic state, 
we obtain transfer integrals and one-particle energy levels. 
We take about 2000 transfer integrals to fit 17 bands near the Fermi energy 
using the 17 Wannier orbitals (all of the Cu-$d$ and O-$p$ states are included). 
Some typical values of calculated transfer integrals are 
1.346 eV between nearest-neighbor Cu-$d_{x^2-y^2}$ and in-plane O-$p_{x/y}$, 
and 0.623 eV between nearest-neighbor in-plane O-$p_{x/y}$. 
These values quantitatively agree with 1.3 eV and 0.65 eV in Ref.~\ref{Ref:Hybertsen1989}. 
One-particle energy levels are obtained for the Cu-$d$ orbitals as: 
$\varepsilon_{xy} = 6.81$ eV, $\varepsilon_{yz/xz} = 7.12$ eV, 
$\varepsilon_{x^2-y^2} = 7.55$ eV, $\varepsilon_{3z^2-r^2} = 7.56$ eV 
(the Fermi level is about 9.53 eV). 
Only the $d_{x^2-y^2}$ state is partially filled, and the other $d$ orbitals 
are completely filled with electrons. 
Here we consider that these values do not reflect a realistic level scheme 
of the local Cu-$d$ orbitals: for example, 
$\varepsilon_{x^2-y^2} - \varepsilon_{yz/xz}  = 0.43$ eV will be too small. 
To reproduce the $d$-$d$ transition energies which were recently determined 
with great accuracy by $L$-edge RIXS~\cite{Sala2011}, 
we modify the one-particle energy levels of the completely-filled 
Cu-$d$ orbitals in the following way: 
we subtract 1.40 eV from $\varepsilon_{\ell}$ 
for the $t_{2g}$ ($d\varepsilon$) orbitals 
and 1.65 eV for the $d_{3z^2-r^2}$ orbital, and as a result we obtain 
$\varepsilon_{xy} = 5.41$ eV, $\varepsilon_{yz/xz} = 5.72$ eV, 
$\varepsilon_{x^2-y^2} = 7.55$ eV, $\varepsilon_{3z^2-r^2} = 5.91$ eV 
(the Fermi level is about 9.49 eV). 
Hereafter, we redefine $\varepsilon_{\ell}$ by the subtracted one-particle energy. 
In Fig.~\ref{Fig:bandsdos}(a), we show the original electronic structure, 
the fitted bands using the 17 Wannier orbitals, 
and the bands obtained for the modified Cu-$d$ energy levels. 

To describe the antiferromagnetic ground state, 
we apply the HF approximation to the tight-binding Hamiltonian $H_{n.f.}$. 
For the Coulomb integrals $I_{\ell_1,\ell_2;\ell_3,\ell_4}$, we introduce the following notation: 
\begin{eqnarray}
J_{\ell,\ell'} & \equiv & I_{\ell,\ell';\ell',\ell}, \\ 
K_{\ell,\ell'} & \equiv & I_{\ell,\ell';\ell,\ell'}. 
\end{eqnarray}
$J_{\ell,\ell'}$ and $K_{\ell,\ell'}$ are the so-called direct and exchange integrals, respectively. 
We assume spin polarization is induced only in the $d$ orbitals at each Cu site, 
and take mean fields only for the Cu-$d$ electrons. 
The mean-field Hamiltonian for $H_{n.f.}$ is 
\begin{eqnarray}
\label{Eq:HMF}
H_{n.f.}^{HF} &=& \sum_{ii'} \sum_{\ell}^{@\mib{r}_i} \sum_{\ell'}^{@\mib{r}_{i'}} 
\sum_\sigma t_{\ell,\ell'}({\mib r}_i-\mib{r}_{i'}) a_{i \ell \sigma}^{\dag} a_{i' \ell' \sigma} 
+ \sum_i^{{\rm t.m.}} \sum_{\ell}^{@\mib{r}_i} \biggl[ \frac{J_{\ell,\ell}}{2} \langle n_{i\ell} \rangle 
+ \sum_{\ell'(\neq \ell)}^{@\mib{r}_i}  \biggl (J_{\ell,\ell'}- \frac{K_{\ell,\ell'}}{2}  \biggr) 
\langle n_{i\ell'} \rangle \biggr] n_{i\ell} \nonumber\\
&& - \sum_i^{{\rm t.m.}} \sum_{\ell}^{@\mib{r}_i} 
\biggl[ \frac{J_{\ell,\ell}}{2} \langle \mib{m}_{i\ell} \rangle 
+ \sum_{\ell'(\neq \ell)}^{@\mib{r}_i} 
\frac{K_{\ell,\ell'}}{2} \langle \mib{m}_{i\ell'} \rangle \biggr] \cdot \mib{m}_{i\ell} 
- \sum_i^{{\rm t.m.}} \sum_{\ell}^{@\mib{r}_i}  \frac{J_{\ell,\ell}}{4} 
\biggl( \langle n_{i\ell} \rangle^2 - | \langle \mib{m}_{i\ell} \rangle |^2 \biggr) \nonumber \\
&& - \sum_i^{{\rm t.m.}} \sum_{\ell \neq \ell'}^{@\mib{r}_i} 
\frac{J_{\ell,\ell'}}{2} \langle n_{i\ell} \rangle \langle n_{i\ell'} \rangle 
+ \sum_i^{{\rm t.m.}} \sum_{\ell \neq \ell'}^{@\mib{r}_i}  \frac{K_{\ell,\ell'}}{4} 
\biggl( \langle n_{i\ell} \rangle \langle n_{i\ell'} \rangle 
+ \langle \mib{m}_{i\ell} \rangle \cdot \langle \mib{m}_{i\ell'} \rangle \biggr), 
\end{eqnarray}
where 
\begin{eqnarray}
n_{i\ell} &=& \sum_{\sigma} d_{i\ell\sigma}^{\dag} d_{i\ell\sigma}, \\
\mib{m}_{i\ell} &=& \sum_{\sigma\sigma'} d_{i\ell\sigma}^{\dag} 
\mib{\sigma}_{\sigma\sigma'} d_{i\ell\sigma'}, 
\end{eqnarray}
using the Pauli matrix vector $\mib{\sigma}$. 
Within the HF theory, we should consider that the one-particle energy $\varepsilon_{\ell}$ 
is already including the following energy shift from the bare one, 
\begin{equation}
\Delta^{HF}_{\ell} \equiv \frac{J_{\ell,\ell}}{2} \langle n_{i\ell} \rangle 
+ \sum_{\ell'(\neq \ell)}^{@\mib{r}_i}  \biggl (J_{\ell,\ell'}- \frac{K_{\ell,\ell'}}{2}  \biggr) 
\langle n_{i\ell'} \rangle, 
\label{Eq:Delta}
\end{equation}
due to the electron-electron Coulomb interaction at transition-metal site $\mib{r}_i$. 
Therefore, before determining the magnetic ground state, 
we need to evaluate the bare one-particle energy by 
$\varepsilon_{\ell}^{(0)} \equiv \varepsilon_{\ell} - \Delta^{HF}_{\ell}$, 
where $\Delta^{HF}_{\ell}$ is evaluated from the expectation values 
of particle numbers $\langle n_{i\ell} \rangle$'s in the paramagnetic state using eq.~(\ref{Eq:Delta}). 
Note that $\varepsilon_{\ell}$'s here are the modified values mentioned above. 
For the Coulomb integrals given in \S~\ref{Sc:Formulation}, 
the obtained values of $\varepsilon_{\ell}^{(0)}$ are as follows: 
$\varepsilon_{xy}^{(0)} = -39.69$ eV, $\varepsilon_{yz/xz}^{(0)} = -40.02$ eV, 
$\varepsilon_{x^2-y^2}^{(0)} = -38.86$ eV, $\varepsilon_{3z^2-r^2}^{(0)} = -40.06$ eV. 
Maintaining these values of $\varepsilon_{\ell}^{(0)}$, we determine the mean-fields 
$\langle n_{i\ell} \rangle$ and $\langle \mib{m}_{i\ell} \rangle$ self-consistently. 
We assume the antiferromagnetic ordering with the spin moments pointing along [110], 
as observed in neutron scattering~\cite{Vaknin1987}. 
As a result, we obtain 68 diagonalized energy bands 
($E_a(\mib{k})$, $1 \leq a \leq 68$) for the antiferromagnetic ground state 
(Note spin degeneracy and folding of the Brillouin zone). 
The $d_{x^2-y^2}$ orbitals accommodate 1.27 electrons and take the spin moment 
of 0.69 $\mu_B$ per Cu site, while the other Cu-$d$ orbitals are completely filled with electrons. 
Figure~\ref{Fig:bandsdos}(b) shows the obtained density of states 
for the antiferromagnetic ground state. 

\section{Resonant X-ray Absorption Spectrum}
\label{Ap:XAS}

\begin{figure}
\begin{center}
\includegraphics[width=90mm]{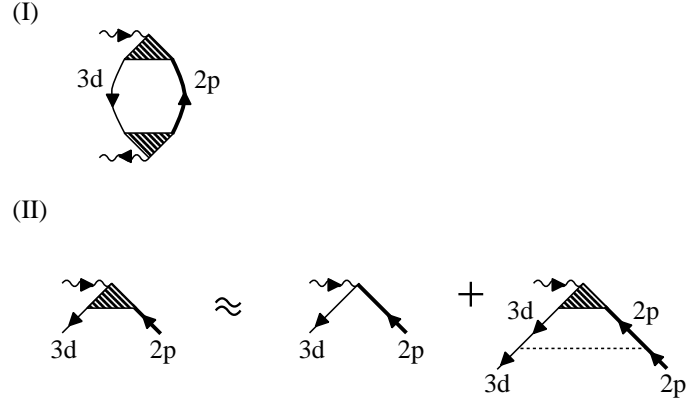}
\end{center}
\caption{
(I) Diagrammatic expression of x-ray absorption intensity. 
Wavy line, thick solid line, and thin solid line represent the propagators for the photon, 
Cu-$2p$ and Cu-$3d$ electrons, respectively. 
The shaded triangle is the electron-photon interaction renormalized 
by the Coulomb interaction between the $2p$ and $3d$ electrons. 
(II) The vertex correction by the multiple scattering 
between the $2p$ and $3d$ electrons (Ladder approximation). }
\label{Fig:diagramsa}
\end{figure}

Resonant x-ray absorption spectrum (XAS) is obtained by calculating the number of photons 
absorbed by the electronic system per unit time. 
Keldysh diagrammatic representation of XAS is presented in Fig.~\ref{Fig:diagramsa}(I), 
and the analytic expression is 
\begin{eqnarray}
A(q, \mib{e}) &=& \frac{1}{N} \sum_{\mib{k}} \int_{-\infty}^{\infty} d t \,
e^{-i \omega t} \sum_i^{\rm t.m.u.} \sum_{\zeta\zeta'}^{@\mib{r}_i} \sum_J 
\sum_{M=-J}^{J} \nonumber \\ 
&& \bar{w}_{\zeta, JM}(\mib{r}_i; q, \mib{e}) 
\bar{w}_{\zeta', JM}^*(\mib{r}_i; q, \mib{e}) 
G_{2p_{JM}(i)}^+(\mib{k}, t) G_{\zeta', \zeta}^-(\mib{k}+\mib{q}, -t),  
\label{Eq:XAS}
\end{eqnarray}
where $q=(\omega, \mib{q})$ and $G_{2p_{JM}(i)}^+(\mib{k}, t)$ 
and $G_{\zeta', \zeta}^-(\mib{k}+\mib{q}, t)$ 
are the Keldysh Green's functions for the $2p_{JM}$ electrons 
at transition-metal site $i$ and that for the transition-metal $d$ electrons, respectively: 
\begin{eqnarray}
G_{2p_{JM}(i)}^+(\mib{k}, t) &=& i n_{2p_{JM}(i)} e^{-i \varepsilon_{2p_J}(\mib{r}_i) t } 
= i e^{-i \varepsilon_{2p_J}(\mib{r}_i) t } \\
G_{\zeta', \zeta}^-(\mib{k}, t) &=&  -i \sum_a u_{\zeta', a}(\mib{k}) u_{\zeta, a}^*(\mib{k}) 
[1-n_a(\mib{k})] e^{i E_a(\mib{k}) t }. 
\end{eqnarray}
Thus eq.~(\ref{Eq:XAS}) is reduced to 
\begin{eqnarray}
A(q, \mib{e}) &=& 2 \pi \sum_i^{\rm t.m.u.} \sum_{\zeta\zeta'}^{@\mib{r}_i} \sum_J 
\sum_{M=-J}^{J} \bar{w}_{\zeta, JM}(\mib{r}_i; q, \mib{e}) 
\bar{w}_{\zeta', JM}^*(\mib{r}_i; q, \mib{e}) \nonumber\\
&& \times \frac{1}{N} \sum_{\mib{k}} \sum_a [1-n_a(\mib{k})] 
u_{\zeta', a}(\mib{k}) u_{\zeta, a}^*(\mib{k}) 
\delta(\omega + \varepsilon_{2p_J}(\mib{r}_i) - E_a(\mib{k})). 
\label{Eq:XAS2}
\end{eqnarray}
$\bar{w}_{\zeta, JM}(\mib{r}_i; q, \mib{e})$ is the effective 2$p$-$d$ 
dipole-transition matrix renormalized within the ladder approximation 
as in Fig.~\ref{Fig:diagramsa}(II): 
\begin{eqnarray}
\bar{w}_{\zeta, JM}(\mib{r}_i; q, \mib{e}) &=& w_{\zeta, JM}(\mib{r}_i; \mib{q}, \mib{e}) 
+ i \sum_{\zeta'\zeta''}^{@\mib{r}_i} \sum_{J'} \sum_{M'=-J'}^{J'} 
V_{2p-d}( \mib{r}_i; J'M', \zeta; \zeta'', JM) \nonumber\\
&& \times \bar{w}_{\zeta', J'M'}(\mib{r}_i; q, \mib{e}) 
\frac{1}{N} \sum_{\mib{k}} \int_{-\infty}^{+\infty} \frac{d z}{2 \pi} \,
G_{2p_{J'M'}(i)}(\mib{k}, z) G_{\zeta'', \zeta'}(\mib{k}+\mib{q}, z + \omega) \nonumber\\
&=& w_{\zeta, JM}(\mib{r}_i; \mib{q}, \mib{e}) 
- \sum_{\zeta'\zeta''}^{@\mib{r}_i} \sum_{J'} \sum_{M'=-J'}^{J'} 
V_{2p-d}( \mib{r}_i; J'M', \zeta; \zeta'', JM) \nonumber\\
&& \times \bar{w}_{\zeta', J'M'}(\mib{r}_i; q, \mib{e}) 
\frac{1}{N} \sum_{\mib{k}} \sum_a u_{\zeta'', a}(\mib{k}) u_{\zeta', a}^*(\mib{k}) 
\frac{1-n_a(\mib{k})}{\omega + \varepsilon_{2p_{J'}}(\mib{r}_i) -E_a(\mib{k}) 
+ i \epsilon}, \nonumber\\ 
\end{eqnarray}
where $G_{2p_{JM}(i)}(k)$ and $G_{\zeta', \zeta}(k)$ are the ordinary causal Green's functions 
for the $2p_{JM}$ electrons at transition-metal site $i$ 
and that for the transition-metal $d$ electrons, respectively. 

\begin{figure}
\begin{center}
\includegraphics[width=90mm]{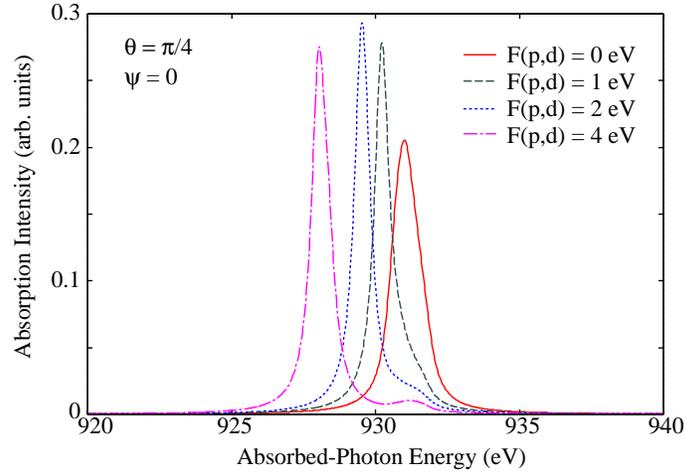}
\end{center}
\caption{(Color online)
Calculated x-ray absorption spectra (XAS) near the Cu $L_3$-edge. 
We show four cases of the Coulomb interaction 
between the Cu-$2p$ and Cu-$3d$ orbitals: $F^0(p,d) = F^2(p,d) = 0, 1, 2, 4$ eV. 
The Bragg angle is $\theta=\pi/4$, and the absorbed photon 
is in $\pi$ polarization ($\psi=0$). }
\label{Fig:rxas}
\end{figure}

Calculated results of XAS for several values of 
the Coulomb interaction between the Cu-$2p$ and Cu-$3d$ orbitals 
are displayed in Fig.~\ref{Fig:rxas}. The main peak position shifts 
to the low-energy region, increasing the Coulomb interaction. 
The calculated spectral shape suggests a possibility that 
a small satellite peak can appear. 
For the calculations of RIXS spectra, we simply use the unrenormalized 
matrix elements $w_{\zeta, JM}(\mib{r}_i; \mib{q}, \mib{e})$. 
The photon energy for the main resonance absorption is changed 
by using renormalized $\bar{w}_{\zeta, JM}(\mib{r}_i; q, \mib{e})$, 
but it can be re-adjusted to the observed value 
by shifting the inner-shell $2p$ level.


\begin{thebibliography}{99}
\bibitem{Ament2011}
L.J.P. Ament, M. van Veenendaal, T.P. Devereaux, 
J.P. Hill, and J. van den Brink, Rev. Mod. Phys. {\bf 83}, 705 (2011). 

\bibitem{Ishii2013}
K. Ishii, T. Tohyama, and J. Mizuki, J. Phys. Soc. Jpn. {\bf 82}, 021015 (2013). 

\bibitem{Hill1998}
J.P. Hill, C.C. Kao, W.A.L. Caliebe, M. Matsubara, A. Kotani, 
J.L. Peng, and L. Greene, 
Phys. Rev. Lett. {\bf 80}, 4967 (1998). 

\bibitem{Kim2002}
Y.J. Kim, J.P. Hill, C.A. Burns, S. Wakimoto, R.J. Birgeneau, D. Casa, 
T. Gog, and C.T. Venkataraman, 
Phys. Rev. Lett. {\bf 89}, 177003 (2002).

\bibitem{Ishii2011}
K. Ishii, S. Ishihara, Y. Murakami, K. Ikeuchi, K. Kuzushita, T. Inami, 
K. Ohwada, M. Yoshida, I. Jarrige, N. Tatami, S. Niioka, D. Bizen, 
Y. Ando, J. Mizuki, S. Maekawa, and Y. Endoh, 
Phys. Rev. B {\bf 83}, 241101 (2011).

\bibitem{Hill2008}
J.P. Hill, G. Blumberg, Y.-J. Kim, D. S. Ellis, S.Wakimoto, R. J. Birgeneau, 
S. Komiya, Y. Ando, B. Liang, R.L. Greene, D. Casa, and T. Gog, 
Phys. Rev. Lett. {\bf 100}, 097001 (2008). 

\bibitem{DeGroot1998}
F.M.F. de Groot, P. Kuiper, and G.A. Sawatzky, 
Phys. Rev. B {\bf 57}, 14584 (1998).

\bibitem{VanVeenendaal2006}
M. van Veenendaal, 
Phys. Rev. Lett. {\bf 96}, 117404 (2006).

\bibitem{Ament2009}
L.J.P. Ament, G. Ghiringhelli, M. Moretti Sala, L. Braicovich, and J. van den Brink, 
Phys. Rev. Lett. {\bf 103}, 117003 (2009). 

\bibitem{Braicovich2010a}
\label{Ref:Braicovich2010a}
L. Braicovich, J. van den Brink, V. Bisogni, M. Moretti Sala, L.J.P. Ament, N.B. Brookes, 
G.M. De Luca, M. Salluzzo, T. Schmitt, V.N. Strocov, and G. Ghiringhelli, 
Phys. Rev. Lett. {\bf 104}, 077002 (2010). 

\bibitem{Braicovich2010b}
L. Braicovich, M. Moretti Sala, L.J.P. Ament, V. Bisogni, M. Minola, 
G. Balestrino, D. Di Castro, G.M. De Luca, M. Salluzzo, G. Ghiringhelli, 
and J. van den Brink,  
Phys. Rev. B {\bf 81}, 174533 (2010). 

\bibitem{Guarise2010}
M. Guarise, B. Dalla Piazza, M. Moretti Sala, G. Ghiringhelli, L. Braicovich, 
H. Berger, J. N. Hancock, D. van der Marel, T. Schmitt, V. N. Strocov, 
L. J. P. Ament, J. van den Brink, P.-H. Lin, P. Xu, H. M. R{\o}nnow, and M. Grioni, 
Phys. Rev. Lett. {\bf 105}, 157006 (2010). 

\bibitem{Schlappa2012}
J. Schlappa, K. Wohlfeld, K.J. Zhou, M. Mourigal, M.W. Haverkort, V.N. Strocov, L. Hozoi, 
C. Monney, S. Nishimoto, S. Singh, A. Revcolevschi, J.-S. Caux, L. Patthey, H.M. R{\o}nnow, 
J. van den Brink, and T. Schmitt, 
Nature {\bf 485}, 82 (2012). 

\bibitem{Dean2012}
\label{Ref:Dean2012}
M.P.M. Dean, R.S. Springell, C. Monney, K.J. Zhou, J. Pereiro, I. Bo\v{z}ovi\'{c}, B. Dalla Piazza, 
H.M. R{\o}nnow, E. Morenzoni, J. van den Brink, T. Schmitt, and J.P. Hill, 
Nature Materials {\bf 11}, 850 (2012). 

\bibitem{LeTacon2011}
M. Le Tacon, G. Ghiringhelli, J. Chaloupka, M. Moretti Sala, V. Hinkov, M.W. Haverkort,
M. Minola, M. Bakr, K.J. Zhou, S. Blanco-Canosa, C. Monney, Y.T. Song, G. L. Sun, C. T. Lin,
G.M. De Luca, M. Salluzzo, G. Khaliullin, T. Schmitt, L. Braicovich, and B. Keimer, 
Nature Physics {\bf 7}, 725 (2011). 

\bibitem{Jia2014}
C.J. Jia, E.A. Nowadnick, K. Wohlfeld, Y.F. Kung, C.-C. Chen, 
S. Johnston, T. Tohyama, B. Moritz, and T.P. Devereaux, 
Nature Communications {\bf 5}, 3314 (2014). 

\bibitem{Ishii2014}
K. Ishii, M. Fujita, T. Sasaki, M. Minola, G. Dellea, C. Mazzoli, K. Kummer, G. Ghiringhelli,
L. Braicovich, T. Tohyama, K. Tsutsumi, K. Sato, R. Kajimoto, K. Ikeuchi, K. Yamada,
M. Yoshida, M. Kurooka, and J. Mizuki, 
Nature Communications {\bf 5}, 3714 (2014). 

\bibitem{Lee2014}
W.S. Lee, J.J. Lee, E.A. Nowadnick, S. Gerber, W. Tabis, S.W. Huang, V.N. Strocov,
E.M. Motoyama, G. Yu, B. Moritz, H.Y. Huang, R.P. Wang, Y.B. Huang, W.B. Wu, 
C.T. Chen, D.J. Huang, M. Greven, T. Schmitt, Z.X. Shen, and T.P. Devereaux, 
Nature Physics {\bf 10}, 883 (2014). 

\bibitem{Tanaka1993}
S. Tanaka and A. Kotani, 
J. Phys. Soc. Jpn. {\bf 62}, 464 (1993).

\bibitem{Tsutsui2006}
K. Tsutsui, H. Yamamoto, T. Tohyama, and S. Maekawa, 
J. Phys. Chem. Solids {\bf 67}, 274 (2006).

\bibitem{Haverkort2010}
M. W. Haverkort, 
Phys. Rev. Lett. {\bf 105}, 167404 (2010).

\bibitem{Kaneshita2011}
E. Kaneshita, K. Tsutsui, and T. Tohyama, 
Phys. Rev. B {\bf 84}, 020511 (2011). 

\bibitem{Igarashi2012}
J. Igarashi and T. Nagao, 
Phys. Rev. B {\bf 85}, 064421 (2012). 

\bibitem{Mostofi2008}
A. A. Mostofi, J. R. Yates, Y.-S. Lee, I. Souza, D. Vanderbilt, and N. Marzari, 
{\tt wannier90}: A Tool for Obtaining Maximally-Localised Wannier Functions,
Comp. Phys. Commun. {\bf 178}, 685 (2008). 

\bibitem{Nomura2005}
T. Nomura and J. Igarashi, 
Phys. Rev. B {\bf 71}, 035110 (2005). 

\bibitem{Nomura2014}
T. Nomura, 
J. Phys. Soc. Jpn. {\bf 83}, 064707 (2014). 

\bibitem{Igarashi2013}
\label{Ref:Igarashi2013}
J. Igarashi and T. Nagao, 
Phys. Rev. B {\bf 88}, 014407 (2013). 

\bibitem{Blaha2013}
P. Blaha, K. Schwarz, G. Madsen, D. Kvasnicka, and J. Luitz, 
WIEN2k, An Augmented Plane Wave Plus Local Orbitals Program
for Calculating Crystal Properties (ISBN 3-9501031-1-2). 

\bibitem{Condon1959}
\label{Ref:Condon1959}
E.U. Condon and G.H. Shortley, {\it The Theory of Atomic Spectra} 
(Cambridge University Press, Cambridge, 1959). 

\bibitem{Czyzyk1994}
\label{Ref:Czyzyk1994}
M.T. Czy\.{z}yk and G.A. Sawatzky, 
Phys. Rev. B {\bf 49}, 14211 (1994).

\bibitem{Ginder1988}
J.M. Ginder, M.G. Roe, Y. Song, R.P. McCall, J.R. Gaines, and E. Ehrenfreund, 
Phys. Rev. B {\bf 37}, 7506 (1988). 

\bibitem{Nozieres1974}
\label{Ref:Nozieres1974}
P. Nozi\`eres and E. Abrahams, 
Phys. Rev. B {\bf 10}, 3099 (1974). 

\bibitem{Keldysh1965}
L.V. Keldysh, 
Sov. Phys. JETP {\bf 20}, 1018 (1965). 

\bibitem{Sala2011}
\label{Ref:Sala2011}
M. Moretti Sala, V. Bisogni, C. Aruta, G. Balestrino, H. Berger, N.B. Brookes, 
G.M. de Luca, D. Di Castro, M., Grioni, M. Guarise, P.G. Medaglia, F. Miletto Granozio, M. Minola, 
P. Perna, M. Radovic, M. Salluzzo, T. Schmitt, K.J. Zhou, L. Braicovich, and G. Ghiringhelli, 
New J. Phys. {\bf 13}, 043026 (2011). 

\bibitem{Igarashi2015}
\label{Ref:Igarashi2015}
J. Igarashi and T. Nagao, 
J. Phys. Condens. Matter {\bf 27}, 186002 (2015).

\bibitem{Braicovich2014}
L. Braicovich, M. Minola, G. Dellea, M. Le Tacon, M. Moretti Sala, C. Morawe, 
J.C. Peffen, R. Supruangnet, F. Yakhou, G. Ghiringhelli, and N.B. Brookes, 
Rev. Sci. Instrum. {\bf 85}, 115104 (2014). 

\bibitem{Hybertsen1989}
\label{Ref:Hybertsen1989}
M.S. Hybertsen, M. Schl\"uter, and N.E. Christensen, 
Phys. Rev. B {\bf 39}, 9028 (1989). 

\bibitem{Vaknin1987}
\label{Ref:Vaknin1987}
D. Vaknin, S.K. Sinha, D.E. Moncton, D.C. Johnston, J.M. Newsam, 
C.R. Safinya, and H.E. King, Jr., 
Phys. Rev. Lett. {\bf 58}, 2802 (1987). 

\end{thebibliography}
\end{document}